\documentclass[11pt]{article}
\pdfoutput=1
\usepackage{jcapmod}
\usepackage{mathtools}
\usepackage{microtype}
\usepackage{ragged2e}
\usepackage[utf8]{inputenc}
\usepackage[english]{babel}
\usepackage[dvipsnames]{xcolor}
\usepackage{amsmath,amssymb,amsbsy,amstext,amsthm}
\usepackage[colorlinks=true]{hyperref}
\usepackage{microtype}
\usepackage{graphicx}
\usepackage{amsfonts}
\usepackage{upgreek}
\usepackage{exscale,relsize}
\usepackage[makeroom]{cancel}
\usepackage{soul}
\usepackage{float}
\usepackage{empheq}
\usepackage[most]{tcolorbox}
\usepackage{ifpdf}
\ifpdf
\fi

\usepackage{colortbl}
\definecolor{green2}{cmyk}{0, 1, 0.5, 0}
\definecolor{lightgreen}{cmyk}{0.2, 0, 0.2, 0.2}
\definecolor{dred}{rgb}{0.9,0.2,0.5}
\definecolor{dred2}{cmyk}{0.1,0.7,0.1,0.3}
\definecolor{lightgray2}{cmyk}{0.4,0.4,0,0.8}
\definecolor{black}{cmyk}{1.0,1.0,1.0,1.0}

\AtBeginDocument{
 \hypersetup{
 urlcolor=NavyBlue,
 citecolor=NavyBlue,
 linkcolor=BrickRed,
 }
}

\allowdisplaybreaks[1]

\usepackage{colortbl}

\makeatletter
\newlength{\apb@width}
\newcommand{\autoparbox}[2][c]{\settowidth{\apb@width}{#2}\parbox[#1]{\apb@width}{#2}}

\makeatother

\numberwithin{equation}{section}

\def\beq{\begin{equation}}
\def\eeq{\end{equation}}

\def\bea{\begin{eqnarray}}
\def\eea{\end{eqnarray}}
\def\eg{{\it e.g.~}}

\def\ie{{\it i.e.~}}
\def\d{{\rm d}}

\def\d{{\rm d}}

\def\nn{\nonumber}

\def\Mp{M_{\rm pl}}

\def\fr{\frac}

\def\0{{\boldsymbol 0}}

\def\fr{\frac}

\newcommand{\vbf}[1]{\mathbf{#1}}

\newtcbox{\mymath}[1][]{%
    nobeforeafter, math upper, tcbox raise base,
    enhanced, colframe=gray!30!gray,
    colback=gray!10, boxrule=0.5pt,
    #1}

\usepackage{setspace} 

\begin{document}

\begin{titlepage}

\setcounter{page}{1} \baselineskip=15.5pt \thispagestyle{empty}

\bigskip\

\vspace{1cm}
\begin{center}

{\fontsize{14}{25}\selectfont  
{\bf Vector dark matter, inflation,  \\and  non-minimal couplings with gravity 
}}

\end{center}

\vspace{0.2cm}
\begin{center}
{\fontsize{13}{30}\selectfont Ogan \"Ozsoy
$^{\dagger}$,\,\, Gianmassimo Tasinato$^{\star, \times}$
}
\end{center}
\begin{center}
\vskip 8pt
\textsl{$^\dagger$ Instituto de Física Téorica UAM/CSIC, Calle Nicolás Cabrera 13-15, Cantoblanco,
28049, Madrid, Spain.} 
\vskip 3pt
\textsl{$^\star$  Department of Physics, Swansea University, Swansea, SA2 8PP, United Kingdom.}
\vskip 3pt
\textsl{$^\times$ Dipartimento di Fisica e Astronomia, Universit\`a di Bologna, via Irnerio 46, Bologna, Italy.}
\end{center}

\vspace{1.2cm}

\noindent
\begin{abstract}

 We propose a cosmological dark matter production mechanism in the form of a longitudinal massive vector boson. We build  upon the work \cite{Graham:2015rva} including non-minimal couplings of the massive vector with gravity, developing a   well motivated set-up   from an effective field theory perspective.
  We carefully track the dynamics of vector field in passing from inflation to radiation dominated universe to show that the late time abundance of longitudinal modes -- excited initially by the quantum fluctuations during inflation -- can provide the observed dark matter abundance for sufficiently weak non-minimal coupling and wide range of vector masses $5 \times 10^{-7} \lesssim m\, [{\rm eV}] \lesssim 5 \times 10^{3}$. The final abundance of dark matter depends on two parameter, the vector mass and its non-minimal coupling with gravity.  We discuss  experimental venues to probe this framework,  including the production of a stochastic gravitational wave background. The latter is especially interesting, as the same mechanism that generates dark matter  can potentially lead to the production of  gravitational waves in the LISA frequency band,  through the second-order effects of large dark matter iso-curvature perturbations at small scales. We take a first step in this direction, identifying the potential information that gravitational wave experiments can provide on the parameter space of dark matter within this scenario.   
\end{abstract}
\vspace{0.6cm}
 \end{titlepage}

\tableofcontents
\newpage 

\section{Introduction}
The nature of dark matter
is one of the most pressing open problems in physics. 
 Current observations 
  provide convincing evidence for gravitational interactions between
    dark matter belonging to   a hidden sector,
   and the visible sector
 of   the Standard Model   \cite{Planck:2018vyg,Sofue:2000jx,Bartelmann:1999yn}.
 If  gravity is the only mediator between the hidden and visible sectors, we need some efficient  mechanism to produce dark matter in the early universe. 
  Cosmic
 inflation is able to do so  by means of  the phenomenon of particle
 production in curved space. See e.g. \cite{Ford:1986sy,Chung:1998zb} for classic examples exploiting this possibility. In this work we consider
 the  recent, compelling
 scenario   \cite{Graham:2015rva},  in which
  dark matter corresponds to the longitudinal components
 of a massive vector field created during inflation. The longitudinal vector fluctuations have a distinctive
 kinetic term structure,  and  specific cosmological   behaviour during inflation and radiation domination. 
Their spectrum grows towards small scales  up to a peak, whose position and amplitude  
 depends on the vector mass. From a cosmological viewpoint, the longitudinal vector fields correspond to iso-curvature fluctuations.
 Their size is  small at large cosmic microwave background (CMB) scales, and
 satisfy CMB constraints.    Besides its elegance, the scenario is  compelling since its predictions
 depend on two parameters only: the present-day dark matter
 abundance depends only on the vector mass and on the energy 
 scale of inflation. 
 Various developments and further
 studies of 
 the original scenario \cite{Bastero-Gil:2018uel,Ema:2019yrd,Ahmed:2020fhc,Kolb:2020fwh,Salehian:2020asa,Moroi:2020has,Long:2019lwl,Arvanitaki:2021qlj} have been considered. Additional early cosmological
 phases different from radiation  enlarge the possible ranges for the vector mass \cite{Ahmed:2020fhc,Kolb:2020fwh}, as
 well as additional couplings with particles in the dark sector \cite{Arvanitaki:2021qlj}. The drawback of these extensions 
 is that often the dark matter abundance depends on additional free parameters, making the set-up
 less predictive. 
  
  \smallskip
  
  We propose  a  generalization of \cite{Graham:2015rva}
  with an action including a non-minimal coupling of the vector field with gravity. The coupling
  we consider 
  is quadratic in the vector field and it 
  has dimension four as the vector field strength. It is the only non-minimal
  coupling with
  these characteristics, and which avoids 
   Ostrogradsky instabilities. Hence, from an effective
   field theory point of view, once the Abelian symmetry
   is broken by a vector mass, our scenario is as minimal and well motivated
   as the original  \cite{Graham:2015rva}. Its predictions on the dark matter
   amplitude again depend on two parameters
   only: the vector mass and the strength  of the non-minimal vector coupling to gravity. 
   The cosmological dynamics of vector longitudinal fluctuations
   is more complex than  \cite{Graham:2015rva} for  the presence of gradient instabilities
   generated by the non-minimal couplings. We discuss in detail how to tame
   such instabilities, and obtain a working scenario with testable predictions. 
   
     \smallskip

One  interesting aspect  of dark matter  constituted by
 longitudinal massive vector  fluctuations -- a {\it dark photon} -- is that it
   can be tested in a variety of ways.  From a particle physics perspective, the
  dark photon
 can be tested through its  milli-charged couplings with the Standard Model
 of particle physics (see e.g. \cite{Lanfranchi:2020crw,Fabbrichesi:2020wbt,Caputo:2021eaa} for  recent comprehensive reviews). 
  The dark photon cosmological evolution
  has distinctive properties, see e.g. \cite{McDermott:2019lch,Lee:2020wfn,Bolton:2022hpt,Amin:2022pzv,Shiraishi:2023zda}.
  It can have testable effects when surrounding black holes, or forming
  cosmic strings, see e.g. \cite{Siemonsen:2022ivj,East:2022ppo,LIGOScientific:2021ffg}.
  It can be  probed with accelerometers \cite{Graham:2015ifn} and gravitational wave experiments, see e.g.  \cite{Pierce:2018xmy,Nomura:2019cvc,PPTA:2022eul,Unal:2022ooa,Yu:2023iog}.
  For the range of vector masses allowed by our set-up, we discuss for the first time 
  the possibility of testing this proposal by means of gravitational waves in the LISA frequency band,
 induced at second order  in perturbations by the vector longitudinal modes (see
 e.g. \cite{Domenech:2021ztg} for a general review on the subject).

\smallskip
We start our discussion with sections \ref{sec_setup}, \ref{s2p1}, and \ref{s2p3} elaborating on our set-up in general terms, and the
dynamics of longitudinal vector fluctuations during inflation and radiation domination.
In section \ref{sec_ab} we show that the abundance of dark matter in our scenario
depends on two parameters only. Section \ref{sec_pheno} discusses phenomenological
implications of vector dark matter in our set-up. Four appendixes cover technical aspects of our results.

\section{The set-up: Cosmology with non-minimally coupled massive vector field}
\label{sec_setup}
We study the dynamics of linearized vector fluctuations propagating through a cosmological space-time, controlled by the quadratic vector action \cite{Tasinato:2014eka,Heisenberg:2014rta}

\beq\label{sgf}
S_{\rm GF}= \int \d^4 x { \sqrt{-g}} \,\bigg[-{\frac{1}{4}F^{\mu\nu}F_{\mu\nu}-\frac{1}{2}\,m^2 A_{\mu}A^{\mu}} + \frac16\, {\alpha^2} G_{\mu\nu} A^{\mu} A^{\nu} 
\bigg],
\eeq
where $A_{\mu}$ is a spin-1 vector field,  $F_{\mu\nu} = \nabla_{\mu} A_{\nu} - \nabla_{\nu} A_{\mu}$ its field strength, and $G_{\mu\nu} = R_{\mu\nu} - R\, g_{\mu\nu} /2$ is the Einstein  tensor. 
The vector $A_{\mu}$ belongs to a dark sector independent from the
Standard Model
of particle physics. In our viewpoint, we consider  $A_{\mu}$  a
 {\it dark photon} (see e.g. \cite{Fabbrichesi:2020wbt}
 for a review on the subject), and we assume that the dark photon (and the dark sector in general) has only gravity-induced interactions with Standard Model matter. 
 
 We use eq. \eqref{sgf} to study the generation of longitudinal vector dark matter, extending
the analysis of \cite{Graham:2015rva} by including the effects of the  $\alpha^2$ contributions. The
additional term  enriches the dynamics of vector fluctuations during radiation
domination.      It                                               
enlarges the parameter space of  scenarios leading to the correct dark matter abundance, 
allowing for a   wider range
of vector masses with respect to \cite{Graham:2015rva}. In order to do so, 
 there is no need of  introducing  non-standard cosmological eras, or to make any hypothesis on the reheating process occurring  between the end of inflation and
the onset of radiation domination \cite{Ahmed:2020fhc,Kolb:2020fwh}.  Importantly, the final dark matter
abundance depends only on the parameters $m$ and $\alpha$ appearing
in eq. \eqref{sgf}, with no dependence on the underlying cosmology (not even the scale
of inflation). This makes a dark matter scenario based on the theory  \eqref{sgf} very minimal,
 and hopefully
easier to test since there are only two parameters to constrain.
 
The Abelian symmetry $A_\mu \to A_\mu + \partial_\mu \lambda(x)$ of the field
strength is broken by the vector mass term
proportional to $m$, and by the non-minimal coupling
to gravity proportional to $\alpha^2$ ($\alpha$ being a dimensionless quantity). From
an effective field theory perspective, once the Abelian symmetry is broken by
the vector mass, the contribution proportional to $\alpha^2$ in action \eqref{sgf}
is unavoidable. In fact,  it is normally generated by loop corrections, since it respects the (lack of) symmetries of the system. It has dimension four as the remaining terms in the Lagrangian density, hence its contributions are in principle as important as the ones of the other terms, since it is not suppressed
by a high-energy scale. Furthermore, it is the only  operator quadratic in the vector field  that non-minimally couples the vector
system to gravity,  without introducing Ostrogradsky instabilities \cite{Tasinato:2014eka,Heisenberg:2014rta}. Hence, our set-up is theoretically well motivated, and we can consider the structure of action \eqref{sgf} to be  as minimal as the one of the  original scenario  \cite{Graham:2015rva}. 
We intend to quantitatively determine    the physically interesting 
ranges
  for the parameter $\alpha$
which allow us to realize dark matter in this set-up \footnote{Non-minimal couplings of vector to curvature have been studied
in some detail in \cite{Nelson:2011sf,Arias:2012az,Alonso-Alvarez:2019ixv} discussing dark matter production
in the early universe through misalignment mechanisms.}.

To  study the cosmological evolution of vector fluctuations, we
expand  the action \eqref{sgf} in terms of the vector components $A_{\mu} = (A_0, A_i)$. The resulting
action is quadratic in these quantities.  
The time component $A_0$ is non-dynamical, and can be integrated out. We are left with three dynamical degrees of freedom: two   transverse vector modes $A_\lambda$ ($\lambda = \pm$), and one longitudinal mode $A_L$. The dynamics of transverse and
longitudinal modes decouple around a  Friedmann-Robertson-Walker (FRW)
background, which is  sourced by a perfect fluid parametrized by a constant equation of state $P = w\rho$.

The resulting vector equations of motion in Fourier space read
(see Appendix \ref{AppA}):
\begin{align}\label{ALandAt}
\nn A''_{L} + \left(1 - \frac{3 (1 + w)\,\alpha^2 H^2 }{2 m_{\rm eff}^2}\right) \fr{k^2}{k^2 + a^2 m_{\rm eff}^2}\, 2a H A'_{L}+ \left(1 - \fr{(1+w)\alpha^2 H^2}{m_{\rm eff}^2}\right)\left(k^2 + a^2 m_{\rm eff}^2\right) A_{L} &= 0,\\
A''_\pm + \left(k^2 + a^2 (m^2_{\rm eff} - (w+1)\, \alpha^2 H^2) \right) A_\pm &= 0\,.
\end{align}
Each mode depends on conformal time $\tau$ and co-moving momentum $k = |\vbf{k}|$, 
$H$ is the Hubble rate  evolving as  ${H}' \,=\, - 3 (1 + w)\,H^2\,a/2$. We introduce an effective time dependent mass squared as 
 
\beq
m_{\rm eff}^2 = m^2 + \alpha^2 H^2\,.
\eeq
In writing equations \eqref{ALandAt}, we indicate with primes derivatives along conformal
time. In the course of this work, depending on what is more convenient, we  work
with conformal time $\tau$ or cosmic time $t$, related  by $d \tau = d t/a(t)$.
In the limit of vanishing non-minimal coupling $\alpha \to 0$, the system of equations \eqref{ALandAt} reduces to minimally coupled Proca set-up studied in \cite{Graham:2015rva}. But, for $\alpha\neq 0$, 
the non-minimal couplings with gravity can  affect
the vector longitudinal mode dynamics considerably.

In what follows, we neglect the back-reaction of vector fields
on the geometry to analyze the vector evolution
 during inflation ($w \simeq -1$) and the subsequent radiation dominated universe (RDU) ($w = 1/3$).  We assume that RDU starts after an efficient, nearly instantaneous reheating process.  During inflation the longitudinal vector dynamics is qualitatively very similar to the original scenario \cite{Graham:2015rva}, leading to an identical scale dependence of the longitudinal power spectrum spectrum that however exhibits a quantitatively different amplitude due to non-minimal coupling. On the other hand, a novel vector dynamics can appear at sub-horizon scales during RDU. In fact, due to the contribution proportional to $\alpha^2$ which multiplies the $k^2$  
 in the first of equation of \eqref{ALandAt}, the longitudinal mode develops a gradient instability, lasting a short period of time. Although brief, this phase is sufficient to potentially affect the vector evolution, and
 needs to be handled with care.  We will do so in what comes next (see section \ref{sec_phate}), finding
 conditions to avoid dangerous instabilities. 

\begin{figure}
\centering
\includegraphics[scale=.5]{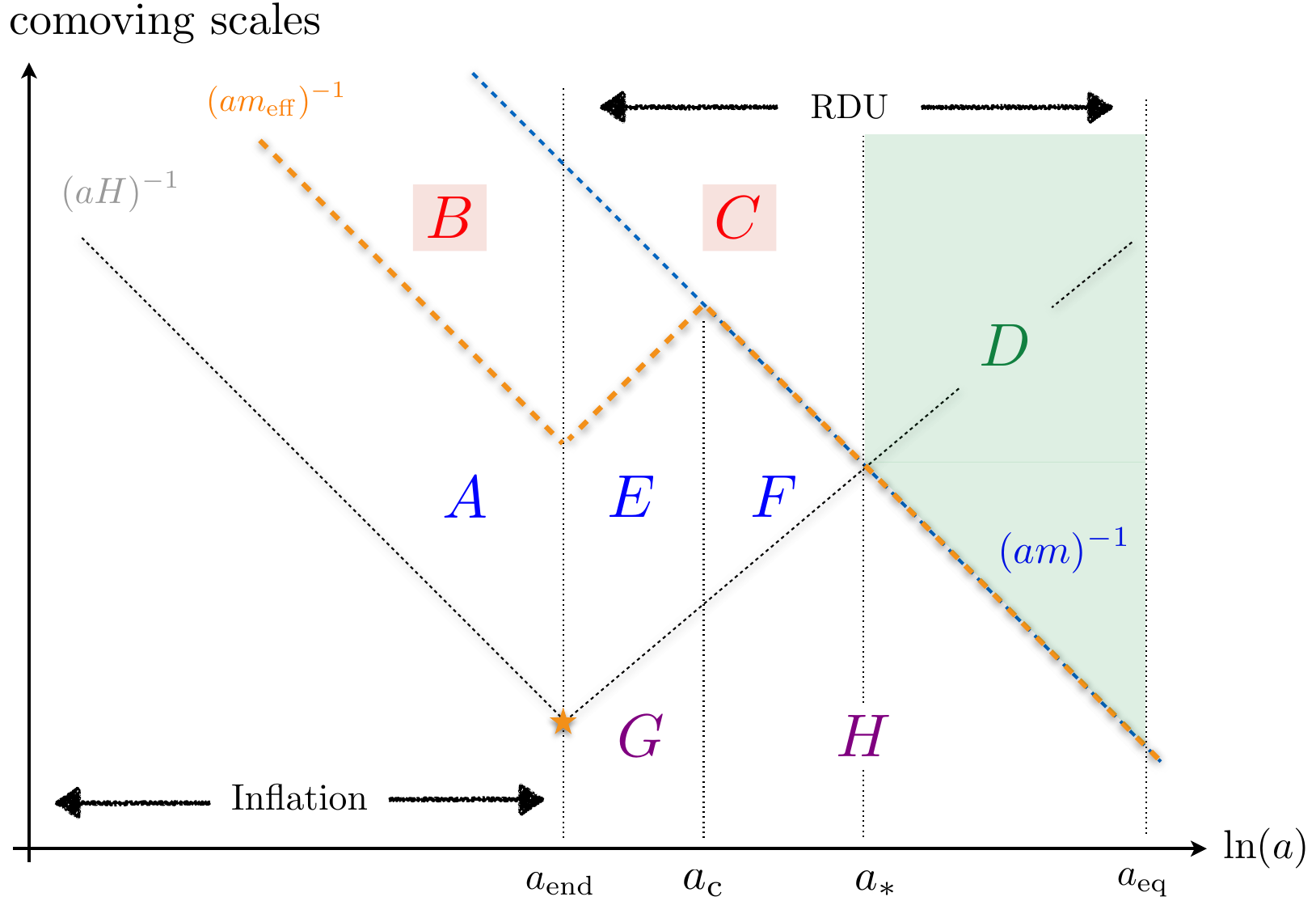}
\caption{A schematic diagram that summarizes the evolution of the longitudinal mode $A_L$ at different stages of the cosmic history. In the main text we will repeatedly 
refer to this figure, explaining the dynamics in each phase denoted by a capital letter.} 
\label{fig:SAL}
\end{figure}

 The resulting cosmological dynamics is  rich, and depends on the cosmological
 epoch one considers, as well as on the size of the vector co-moving wave-number $k$ with respect
 to the remaining parameters of the system. We visually represent the various cosmological phases  in Fig \ref{fig:SAL}.  
   Each epoch is denoted by a capital letter in boldface. It has  different consequences
   for the longitudinal vector evolution, which we study 
 in the  next sections. We make use
 both numerical and analytical tools, inspired from the important papers  \cite{Graham:2015rva,Ema:2019yrd,Ahmed:2020fhc,Kolb:2020fwh}.
 The results will then be used to investigate the properties of dark matter consisting of longitudinal vector modes associated with the theory of eq. \eqref{sgf}. 
 
\section{Dynamics of a light vector field during inflation}
\label{s2p1}

As we discuss below, the vector dynamics during inflation is qualitatively similar to the results in  \cite{Graham:2015rva}. However, for the range of parameters on which we are interested, we will find that the amplitude of the vector power spectrum is considerably affected by the non-minimal
coupling to gravity. To study the dynamics of $A_L$ and $A_{\pm}$ during inflation, we neglect sub-leading slow-roll corrections. We work with a constant Hubble rate $H \simeq H_{\rm I}$, and an equation of state parameter $w = -1$. The scale factor is given by $a(\tau) = 1/(-H_{\rm I}\,\tau)$ in term of the (negative) conformal time $-\infty < \tau \leq 0$
during inflation. The equations of motion (EoM) satisfied by longitudinal and transverse modes  
are
\begin{align}\label{ALandAtinf}
\nn A''_{L} +  \fr{k^2}{k^2 + a^2 m_{\rm eff,I}^2}\, 2a H A'_{L}+ \left(k^2 + a^2 m_{\rm eff,I}^2\right) A_{L} &= 0,\\
A''_\pm + \left(k^2 + a^2 m_{\rm eff,I}^2 \right) A_\pm &= 0\,.
\end{align}
The effective mass reads
\beq
\label{defem1}
m_{\rm eff, I}^2 = m^2 + \alpha^2 H_{\rm I}^2\,,
\eeq
and is constant during inflation.  The structure of these equations 
is identical to the one studied in \cite{Graham:2015rva}. Only the value \eqref{defem1} of the effective
mass $m_{\rm eff, I}$ changes, since it receives a contribution  proportional
to $\alpha^2$ due to  the non-minimal coupling with gravity.

\smallskip

We focus on a region in parameter 
space where the effects of the non-minimal
coupling with gravity is sizeable.  This will make a difference with respect to 
\cite{Graham:2015rva}. For this reason, we
satisfy the inequalities
\beq
\label{ines1a}
m^2\ll \alpha^2 H_{\rm I}^2 \hskip0.7cm,\hskip0.7cm\alpha^2\ll1\,.
\eeq
The first inequality ensures that the non-minimal couplings to gravity, controlled by $\alpha$,  plays an important
role in our discussion. Furthermore, the
second inequality places us in a light-vector regime during inflation,  characterized by 
 $m_{\rm eff, I} \ll H_{\rm I}$. Hence the co-moving Hubble horizon $(a H_{\rm I})^{-1}$ is always much smaller than the   Compton co-moving horizon $(a m_{\rm eff, I})^{-1}$ of the vector field during inflation. In this regime, individual modes of the vector  are guaranteed to satisfy  the relativistic condition $k \gg am_{\rm eff, I}$ during inflation, at least until the time of horizon crossing where $k = a H_{\rm I}$. This condition is  particularly important for understanding the time evolution of the longitudinal mode $A_L$. 

\subsection{Dynamics of the transverse modes}
We
 first focus on the transverse vector modes. We can be brief since their dynamics is the standard one
 of vector fields during inflation.
  In a de Sitter limit, the solution to the mode functions  reducing
   to the standard Bunch Davies vacuum deep inside the horizon gives
\beq
{A}_\pm = \frac{1}{\sqrt{2 k}}\,{\rm e}^{{i(2 \nu+1) \pi}/{4}} \, \sqrt{\frac{\pi x}{2}} H_\nu^{(1)}(x), \quad\quad \nu^2 \equiv \frac{1}{4}-\frac{m_{\rm eff,I}^2}{H_{\rm I}^2},
\eeq
where $H^{(1)}_\nu$ is the Hankel function of first kind and we defined $x = -k\tau$. In the light vector field regime $m_{\rm eff,I}/H_{\rm I} \to 0$
we have $\nu \simeq 1/2$, $H^{(1)}_{1/2} (x) \propto e^{ix}/\sqrt{x}$ and so the transverse modes remain in their vacuum configuration as they are swept outside the horizon during inflation. This is expected since their equation of motion (EoM) \eqref{ALandAtinf} obeys the same equation of a scalar field conformally coupled to gravity in the small effective mass limit. Hence  a sizeable contribution to the transverse mode population, if any, can only be due  to some dynamics after the end of  inflation. We  address this topic in Appendix \ref{sec_trab}, where we show that the post-inflationary dynamics does not manage to appreciably 
increase the amplitude of transverse vector fields.

\subsection{Dynamics of the longitudinal modes}
\label{sec_lmd}
\begin{figure}
\centering
\includegraphics[scale=.8]{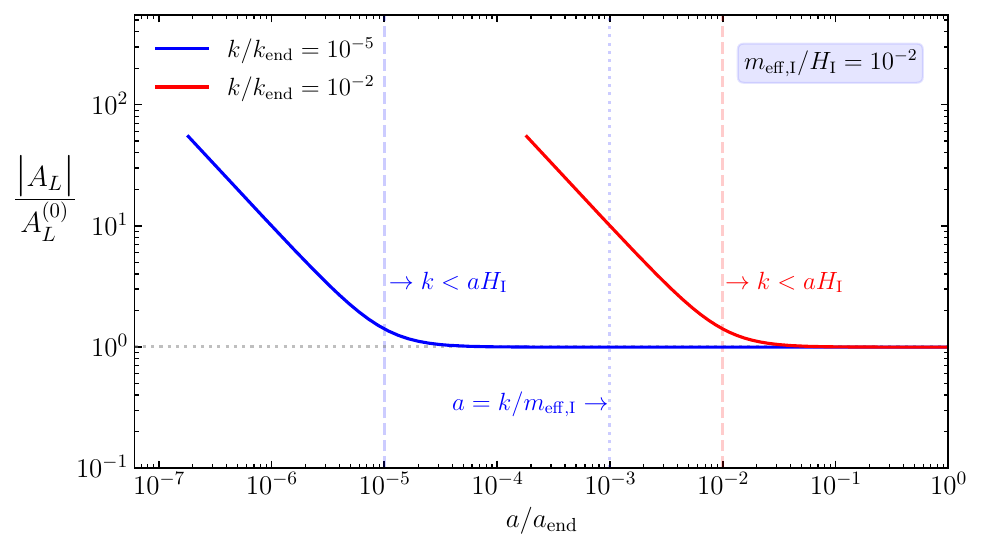}
\caption{Time evolution of two longitudinal modes for $m_{\rm eff, I}/H_{\rm I} = 10^{-2}$ from 4-folds before the horizon exit until the end of inflation where $a/a_{\rm end} = 1$. The dashed vertical lines denote the time of horizon crossing and the dotted vertical line refers to the time when the corresponding mode becomes non-relativistic. Note that the small scale mode labeled by red color stays relativistic throughout its evolution.\label{fig:AL}}
\end{figure}

The dynamics of longitudinal modes is definitely more interesting. 
 To understand the evolution of $A_L$ during inflation, we first notice that -- due to the hierarchy $m_{\rm eff, I} \ll H_{\rm I}$ --  all modes start their evolution in the relativistic regime $k \gg a\, m_{\rm eff,I}$, where they remain  until after horizon crossing. The dynamics of the longitudinal mode is then equivalent to a massless scalar field, up to an overall (important!) re-scaling. In fact, we define
 the canonical variable 
 \beq
 \label{fire1}
 Q_L = (a m_{\rm eff, I}/k) A_L\,,
 \eeq
We  rewrite the EoM \eqref{ALandAtinf}  as
\beq\label{EQ}
Q''_L + \left(k^2 - \fr{a''(\tau)}{a(\tau)}\right)Q_L = 0\,,
\eeq
in a $k \gg a m_{\rm eff,I}$ limit. 
During inflation $a''/a = 2/\tau^2$; consequently,  the solution of eq  \eqref{EQ}   matching the Bunch Davies vacuum deep inside the horizon ($- k \tau \gg 1$) is  
\beq
Q_L = \fr{a H_I}{\sqrt{2\,k^3}}\,\left(1 -   \frac{i k}{a H_{\rm I}}\right)\, {e^{  {i k}/({a H_{\rm I}})}}\,.
\eeq
Inverting the field redefinition \eqref{fire1} by writing $A_L = (k/am_{\rm eff,I}) Q_L$, the solution for the longitudinal mode in the relativistic regime is
\beq\label{LRS}
A_L = A^{(0)}_L(k)\, \,\left(1 -   \frac{i k}{a H_{\rm I}}\right)\, {e^{  {i k}/({a H_{\rm I}})}}\,,
\eeq
where we defined the  $k$-dependent super-horizon amplitude  as
\beq\label{LRSA}
A^{(0)}_L(k) = \fr{1}{\sqrt{2k}} \fr{H_{\rm I}}{m_{\rm eff, I}}.
\eeq
Utilizing \eqref{LRS} and \eqref{LRSA}, we numerically solve the full mode equation \eqref{ALandAtinf} during inflation, by initializing each mode 4 e-folds before they exit the horizon. An example of evolution for two different modes initialized in the relativistic regime is shown in Figure \ref{fig:AL}. 
Shortly after horizon exit, all modes settles into a constant amplitude given by \eqref{LRSA}, 
irrespective of a given mode becoming non-relativistic or not.
We represent this behaviour in
   the red curve of Figure \ref{fig:AL}, which shows the dynamics of a mode that never becomes non-relativistic during inflation. Hence, all longitudinal modes freeze out quickly on super-horizon scales, and settle to a constant amplitude given by
\beq\label{amp}
|A_L |\xrightarrow{ x\, \to \, 0} \fr{1}{\sqrt{2k}}\, \fr{H_{\rm I}}{m_{\rm eff, I}} \equiv A^{(0)}_L(k), \quad\quad\quad |A_L'| \xrightarrow{ x\, \to \, 0} 0\,.
\eeq
The behaviour of $A_L$ we described  can be also understood analytically,  by studying the mode evolution in a piece-wise manner during inflation \cite{Graham:2015rva,Ahmed:2020fhc,Kolb:2020fwh}. To do so,  it is convenient to pass from conformal
to standard time, $d \tau\,=\, d t/a(t)$. 
Recall that $m_{\rm eff}^2 = m_{\rm eff, I}^2 \simeq const.$, $\partial_t m_{\rm eff} = 0 $ and $ w = -1$ during inflation. The EoM \eqref{ALandAtCT} of the longitudinal modes, when
expressed in terms of cosmic time,   reads
\beq\label{ALCTI}
\ddot{A}_L + \frac{3k^2 + a^2 m_{\rm eff, I}^2}{k^2 + a^2 m_{\rm eff, I}^2}\, H\dot{A}_L
 + \left(\frac{k^2}{a^2} + m_{\rm eff,I}^2\right)A_L = 0
\eeq
We now focus on the super-horizon relativistic and non-relativistic regimes, by taking the corresponding limits of \eqref{ALCTI}. We follow the analytic methods developed in \cite{Graham:2015rva} and in \cite{Ahmed:2020fhc,Kolb:2020fwh}. 
From now on, capital letters in boldface refer to different evolution phases
as represented in Fig \ref{fig:SAL}.

\subsubsection{Phase (A): Super-horizon relativistic regime, $H_{\rm I} \gg k/a \gg m_{\rm eff, I}$}

\noindent
 In the relativistic regime the solution to the longitudinal mode is given by \eqref{LRS}. Shortly after horizon exit this solution can be approximated as 
\beq
A_L \simeq A^{(0)}_L(k)\left(1 + \fr{k^2}{2a^2 H_{\rm I}^2}\right)\,,
\eeq
which leads to a function containing a (growing) constant mode, plus  a decaying mode $\propto a^{-2}$:
\beq\label{A}
A_L = c_1 + c_2\, a^{-2},\quad\quad \textrm{{\bf (A)}}\,.
\eeq

\subsubsection{Phase (B):  Super-horizon non-relativistic regime, $H_{\rm I}> m_{\rm eff, I} \gg k/a$ }

In this limit, we can approximate the EoM as ($ H \d t = \d a / a$):
\beq\label{SHNR}
\fr{\d}{\d a}\left(a^2 \fr{\d A_L}{\d a}\right) \simeq -\fr{m_{\rm eff, I}^2}{H_{\rm I}^2} A_L\,.
\eeq
We distinguish two situations. 
In the case of  ultralight fields, the mass contribution in the right
hand side can be neglected, ${m_{\rm eff, I}^2}/{H_{\rm I}^2}\,=\,0$.
 The solution is given by
\beq\label{LB}
A_L = c_1 + c_2 \, a^{-1}, \quad\quad  {m_{\rm eff, I}^2}/{H_{\rm I}^2}\to 0,\,\,\,\textrm{{\bf (B)}}.
\eeq
For not-so-light vector fields however, the right hand side of eq. \eqref{SHNR} plays a role.
We generate a particular solution of \eqref{SHNR} using iteratively the growing constant solution derived from the homogeneous equation. This solution has a secular growth suppressed by the square of the ratio $m_{\rm eff}/H_{\rm I}$, which leads to the final solution
\beq\label{SLB}
A_L = c_1 + c_2 \frac{m_{\rm eff, I}^2}{H_{\rm I}^2}\ln a, \quad\quad  m_{\rm eff, I}\ll H_{\rm I},\,\,\,\textrm{{\bf (B)}}.
\eeq
At first sight the secularly growing term might appear as an issue. However, since all the modes that become non-relativistic   evolve through the relativistic regime during inflation, the contribution
proportional to $ \ln a$ {\it never} becomes dominant
against the constant term.
 This conclusion is confirmed by the behavior of the blue curve in Figure \ref{fig:AL} where the mode enters into the non-relativistic regime (at $a = k/m_{\rm eff, I}$) at an intermediate time denoted by vertical dotted lines. Therefore, the only influence of the secular term is on the derivative of longitudinal mode, which changes its behavior from $A'_L(a) \propto a^{-3}$ (see \eg eq. \eqref{A}) to $A'_L(a) \propto a^{-1}$ (see eq. \eqref{SLB}) as the mode evolve from relativistic to non-relativistic regime. However, in the ultra-light mass limit, 
  eq. \eqref{LB} is still valid in the non-relativistic regime and the dynamics of $A_L$ is captured by eqs. \eqref{A} to \eqref{LB} as the modes evolve from relativistic to non-relativistic regime. (There are also modes that never becomes non-relativistic during inflation, the dynamics of them being  described by the solution \eqref{A}.)

\subsection{The  power spectrum of longitudinal modes at the end of inflation}
\label{sec_ps1}

Collecting the previous results, we conclude that 
super-horizon longitudinal vector modes acquire a power spectrum given by
\beq\label{psinf}
\mathcal{P}_{A_L}(\tau, k) \equiv \fr{k^3}{2\pi^2} \big|A_L(\tau,k)\big|^2 \quad \to \quad \mathcal{P}_{A_L}(\tau_{\rm end}, k) = \left(\fr{k H_{\rm I}}{2 \pi m_{\rm eff, I}}\right)^2.
\eeq
We learn, as \cite{Graham:2015rva},  that the power spectrum at the end of inflation is suppressed  at large scales ($k \to 0$) while it  increases as $k^2$ towards  large $k$ (small scales). 
Such a  characteristic slope of the longitudinal  spectrum is one of the main findings of \cite{Graham:2015rva}. Given that  longitudinal vector perturbations act as iso-curvature modes, 
we then find negligible contributions from iso-curvature modes at large cosmic microwave background (CMB) scales. Hence, this scenario is safe from the point of view of CMB constraints on iso-curvature fluctuations. 

Recall that, given the inequality \eqref{ines1a},
we are interested on the regime of small  bare mass
$
m^2\,\ll\, \alpha^2\,H^2_I
$ with respect to the contribution induced by non-minimal couplings to gravity. Consequently, during inflation, $m_{\rm eff, I}^2\,\simeq\,\alpha^2 H_I^2$, and the longitudinal vector spectrum  is given by
\beq
\label{infPSA}
\mathcal{P}_{A_L}(\tau_{\rm end}, k) = \fr{k^2 }{4 \pi^2 \alpha^2}\,.
\eeq
Notice that the amplitude of the power spectrum is  {\it
 independent} from the vector mass and from the Hubble parameter during inflation.
 This remarkable feature, which is associated with  the non-minimal coupling with gravity, makes a difference from the Proca
 case \cite{Graham:2015rva}, and will be important in what follows
 when computing the dark matter abundance.  
Starting from these initial conditions,
the behaviour of the power spectrum at later times depends on the post-inflationary evolution of the fluctuations,
 as they evolve from super-horizon to sub-horizon scales after inflation ends. This is the argument of our  next discussion,
where we explore  the consequences of the non-minimal coupling to gravity for the abundance of vector
dark matter.

\section{Dynamics of the longitudinal vector mode during radiation domination}
\label{s2p3}

During RDU the vector dynamics at small scales  is considerably influenced by the non-minimal coupling with
gravity. As we will learn, the latter
 induces a gradient instability in the longitudinal vector sector, which
 must be handled with care to maintain the system under control. As a consequence,
 to avoid catastrophic instabilities 
  we will find constraints
 on the available parameter space.

\begin{figure}
\centering
\includegraphics[scale=.45]{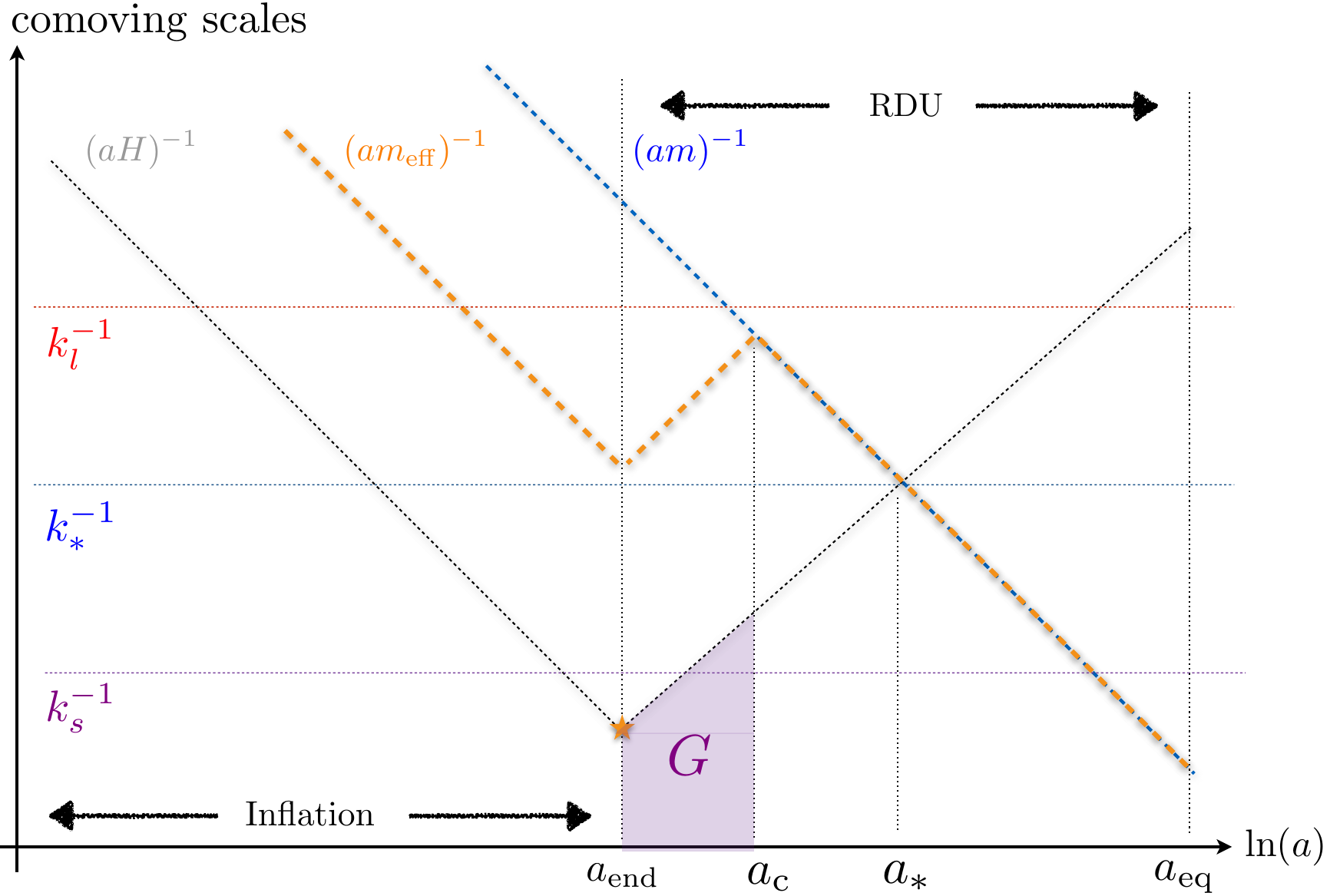}
\caption{The evolution of the individual modes as they evolve from sub-horizon during inflation to super-horizon then back into the horizon in the post-inflationary era.\label{fig:SAL2}}
\end{figure}

In order to understand the effects of the non-minimal coupling $\alpha$ of the vector with gravity, we first study the system analytically. We make 
simplifying assumptions on the time-dependent profiles of the quantities involved, borrowing from previous studies. Our analytical findings will then
be corroborated by a numerical analysis. Since  part
of the dynamics is identical to the Proca case, we can be brief where there
is strong overlap with \cite{Graham:2015rva}. After our analytical
considerations, when  turning to numerics, we focus most of our attention on the  epoch ${\bf (G)}$, which is specific of our set-up, and where the dangerous  gradient instability might develop. See Fig \ref{fig:SAL2}. 

\subsection{Analytical considerations}
\label{sec_anco}

We set $w = 1/3$,  $m_{\rm eff}^2 = m^2 + \alpha^2 H^2$. We recall that
the Hubble parameter during RDU scales as
\beq  \label{evhrd}
H = H_{\rm I} \,\left(\frac{a_{\rm end}}{a}\right)^2\,,
\eeq
where $H_{\rm I}$ is the constant Hubble parameter during inflation, and
$a_{\rm end}$ the scale factor at the end of inflation. In writing eq. \eqref{evhrd}, we
make the hypothesis of instantaneous transition from a (quasi) de Sitter inflationary stage
to RDU.  
The evolution equation for the vector longitudinal mode, when expressed in terms
of cosmic time $t$, is given in eq 
 \eqref{ALandAtCT} and reads
\beq\label{ALRDU}
\ddot{A}_L + \frac{[3+2 \partial_t m_{\rm eff} / (H m_{\rm eff})]\,k^2 + a^2 m_{\rm eff}^2}{k^2 + a^2 m_{\rm eff}^2}\, H\dot{A}_L
 + \left(1 - \frac{4 \alpha^2 H^2}{3 m_{\rm eff}^2}\right)\left(\frac{k^2}{a^2} + m_{\rm eff}^2\right)A_L = 0.
\eeq

Recall that  we are interested in the regime where non-minimal coupling dominates
during inflation, see the inequality \eqref{ines1a}. The hierarchy $\alpha H_I\gg m$ implies that time dependent effective mass is decreasing in the initial stages of RDU, until it reaches its asymptotic value $m$ at a critical time denoted by $a_{\rm c}$:
\beq\label{meff}
m_{\rm eff}^2=
 \begin{dcases} 
       \alpha^2 H_{\rm I}^2 \left(\fr{a_{\rm end}}{a}\right)^4,& \quad a_{\rm end} \leq a \leq a_{\rm c} \,,\\
        m^2 & \quad\quad\quad\quad a > a_{\rm c}\,.
   \end{dcases}
\eeq
The critical scale factor $a_{\rm c}$ sets an important time
for the cosmological evolution of longitudinal modes and is defined
by the relation
\beq\label{DG}
\fr{a_{\rm c}}{a_{\rm end}} \simeq \sqrt{\frac{\alpha H_{\rm I}}{m}} 
\eeq
During a fraction of the cosmic evolution $ a_{\rm end} \leq a \leq a_{\rm c}$, corresponding
to phase ${\bf (G)}$ in Fig \ref{fig:SAL2}, the coefficient of $k^2$
in eq. \eqref{ALRDU} becomes negative, signalling a gradient instability. This
is a new effect of the non-minimal coupling with gravity, that we need to track with care. Notice that the duration of this phase depends on the size non-minimal coupling or more precisely on the ratio $\alpha H_{\rm I}/m$ whose amplitude will be important to keep the fluctuations that re-enter the horizon at this phase under control (see section \ref{sec_phate}).  

Given the time dependence of the effective mass in \eqref{meff}, we  parametrize its
time evolution
in a piece-wise manner as
\beq\label{Dmeff}
\fr{\partial_t m_{\rm eff}}{H m_{\rm eff}} \simeq
 \begin{dcases} 
       -2 ,& \quad a_{\rm end} \leq a \leq a_{\rm c} \,,\\
       \,\,\,\, 0 & \quad\quad\quad\quad a > a_{\rm c}\,,
   \end{dcases}
\eeq
which constitutes
a useful approximation for analytically  handling the friction term in eq  \eqref{ALRDU}.
At the light of the discussion above, we now study the analytic  evolution of the longitudinal modes both for super and sub-horizon scales, during   the different epochs represented in Fig \ref{fig:SAL}.

\subsubsection{Phases (E, F): super-horizon relativistic regime, $H \gg k/a \gg m_{\rm eff}$}

In this regime, the EoM \eqref{ALRDU} is reduced to 
\begin{align}\label{SupHRrdu}
\nn \fr{\d}{\d a} \left(a^2 H \fr{\d A_L}{\d a}\right) - 2 H a \fr{\d A_L}{\d a} &\simeq 0, \quad\quad a_{\rm end} \leq a \leq a_{\rm c}, \quad {\bf (E)}\\
\fr{H}{a^2} \fr{\d}{\d a} \left(a^4 H \fr{\d A_L}{\d a}\right) &\simeq 0, \quad\quad  a > a_{\rm c}, \quad {\bf (F)}\,.
\end{align}
In writing the second line  we assume that the factor  multiplying the mass term in eq. \eqref{ALRDU} quickly reaches an order-one value:
\beq
\left(1 - \frac{4}{3}\frac{\alpha^2 H^2}{m_{\rm eff}^2}\right) \to 1, \quad\quad a > a_{\rm c}.
\eeq
This limit can be justified by re-writing the second term inside the parentheses above as $\alpha^2 H^2/m_{\rm eff}^2 \simeq (a_{\rm c}/a)^4$ for $a > a_{\rm c}$ (recall the scaling \eqref{evhrd}). 
The solutions of eqs \eqref{SupHRrdu} in the super-horizon relativistic regime of RDU are then given by 
\begin{align}\label{SupHRrdus}
\nn A_L & = c_1 + c_2\, a^3,  \quad\quad a_{\rm end} \leq a \leq a_{\rm c}, \quad {\bf (E)}\\ 
A_L &= c_1 + c_2\, a^{-2}, \quad\quad a > a_{\rm c}, \quad \quad {\bf (F)}.
\end{align}
Notice that there is a growing mode
in epoch ${\bf (E)}$. However, the inflationary stage drives the time derivative of the longitudinal mode to extremely small values, and singles out the constant mode to be the dominant one on super-horizon scales. Therefore in the subsequent evolution during RDU, the solution in epoch {\bf (E)}  remains constant on super-horizon scales. We numerically confirmed this behaviour within the parameter space of interest.

\subsubsection{Phases (G, H): sub-horizon relativistic regime, $k/a \gg m_{\rm eff}, H$}
\label{sec_grinan}

Similarly, in the sub-horizon regime, the EoM of longitudinal mode can be reduced to (we switch back to conformal time for convenience)
\begin{align}\label{SHRrdu}
\nn \left(\partial_\tau^2 - 2aH\partial \tau - \fr{k^2}{3}\right)A_L &\simeq 0, \quad\quad a_{\rm end} \leq a \leq a_{\rm c}, \quad {\bf (G)}\\
\left(\partial_\tau^2 + 2aH\partial_\tau + k^2\right) A_L &\simeq 0, \quad\quad  a > a_{\rm c}, \quad {\bf (H)},
\end{align}
 The solutions to these equations can be expressed as
\begin{align}\label{SHRrdus}
\nn A_L &=  c_1 \left[1 - \fr{c_s k}{a H}\right]\, \exp\left(\fr{c_s k}{a H}\right) + c_2 \left[1 + \fr{c_s k}{a H}\right]\, \exp\left(-\fr{c_s k}{a H}\right)\,, \quad\quad a_{\rm end} \leq a \leq a_{\rm c} \quad {\bf (G)}, \\
A_L &= a^{-1} \left( c_1\, e^{ik\tau}+ c_2\, e^{-ik\tau}\right) \quad\quad\quad\quad\quad\quad\quad\quad\quad\quad\quad\quad\quad\quad\quad\quad a > a_{\rm c}, \quad \quad {\bf (H)},
\end{align}
where we defined $c_s = 1/\sqrt{3}$ since  the first line of \eqref{SHRrdu} resembles
the equation for a driven harmonic oscillator with an imaginary sound speed. The first {\it
exponentially growing} solution in \eqref{SHRrdus} is due to the aforementioned  gradient instability for the short scale modes 
within the horizon (phase {\bf (G)})
associated with the non-minimal coupling with gravity. See eq  \eqref{ALRDU} and Fig \ref{fig:SAL2}. 
Given its dangerous exponential amplification, we  need to be sure to keep these short-scale
modes under control. In fact, soon we will learn that  this condition
 imposes limitations on the available parameter space. 
 Notice also that after crossing phase ${\bf (G)}$ the modes
enter in phase ${\bf (H)}$ (see Fig. \ref{fig:SAL}) where their amplitude decreases. This behaviour
will be important for the numerical considerations of Section \ref{sec_phate}.

\subsubsection{Phase (C): Hubble damped non-relativistic regime, $ H \gg  m_{\rm eff} \gg k/a$}

In this regime we neglect terms proportional to $k$ in the damping contributions, as well as the mass term in \eqref{ALRDU} as compared to the damping term. We obtain the evolution equation
\beq
H \fr{\d}{\d a}\left(a^2 H \fr{\d}{\d a} A_L\right) \simeq 0,
\eeq
with  the following solution
\beq
\label{growpC}
A_L = c_1 + c_2\, a, \quad\quad {\bf (C)}.
\eeq
Similar to the analysis we made for epoch ${\bf (F)}$, the growing mode in eq. \eqref{growpC}  does not take over, since the initial conditions inherited from the  inflationary stage singles out the
constant mode as the dominant one.

\subsubsection{Phase (D):  Late time non-relativistic regime, $ m_{\rm eff} \gg k/a, H$}\label{NRAL}

In this regime, we can again neglect $k^2$ contributions in the damping term and  the mass term, see \eqref{ALRDU}: 
\beq
\left(\partial_t^2 + H \partial_t + m^2\right)A_L \simeq 0\,.
\eeq
The solution of this equation is
\beq\label{ALNR}
A_L = a^{-1/2} (c_1\, e^{i m t} + c_2\, e^{- i m t}).
\eeq
As the modes enter the non-relativistic regime, it starts oscillating  with a decaying envelope $a^{-1/2}$  (see Fig. \ref{fig:SAL} and \ref{fig:SAL2}). As first noticed in \cite{Graham:2015rva}, for light vector fields, the energy density contained in the longitudinal mode starts to act like a dark matter in this phase.

\subsection{Numerical Analysis}\label{Numrdu}

The analytical considerations of section \ref{sec_anco} show that during RDU the longitudinal mode dynamics is quite rich and diverse. 
We now need a more careful analysis and a numerical treatment to follow the evolution
of modes from inflation deep into RDU, taking into account the transitions
among different epochs,  and  examining  the
role of the exponential gradient instability we anticipated in section \ref{sec_grinan}. We intend
to explore a region in parameter space where the gradient instability can be tamed, and
where at the same time non-minimal 
couplings to gravity lead to  sizeable, interesting effects for
the phenomenology of  dark matter.

For this purpose, we  follow the cosmological evolution of a dimensionless quantity $T_L$ playing the role of a transfer
function:
\beq\label{Alrdu}
A_L(\tau,k) \,\equiv\, T_L (k\tau)\, A^{(0)}_L(k),
\eeq

where the initial condition $A^{(0)}_L$ is given  in \eqref{amp} in terms
of the  value of the longitudinal mode  at the onset of inflation: see section
\ref{sec_lmd}. To determine accurate initial conditions for the transfer function at the beginning of RDU, we  evolve $T_   L$ for a given mode $k$,  from the time of its  horizon crossing until the end of inflation. We use the numerical procedure described in Appendix \ref{AppB}. This method allows
us to determine
  the initial conditions at the beginning of RDU. We  use them to follow  the mode evolution during RDU, to
finally find the longitudinal vector power spectrum at late times. We express the latter as
\beq\label{LPS}
\mathcal{P}_{A_L} (\tau,k) = \fr{k^3}{2\pi^2}\Big|A^{(0)}_L\Big|^2 \big|T_L (k\tau)\big|^2 = \left(\fr{k_* H_I}{2 \pi m_{\rm eff, I}}\right)^2  \left(\fr{k}{k_*}\right)^2\, \big|T_L (k\tau)\big|^2,
\eeq
in terms of the transfer function $T_L$.
We introduce the quantity $k_*$ corresponding to  the wave-number that re-enters the horizon at the time when the co-moving Hubble horizon is equal to the Compton horizon, as set by the bare mass (see Figs.
 \ref{fig:SAL} and  \ref{fig:SAL2}): 
\beq\label{cs}
(a_* m)^{-1} = (a_* H_*)^{-1} = \tau_* \quad\Longrightarrow\quad \tau_* \equiv k_*^{-1} = \fr{1}{a_* m}. 
\eeq
\begin{figure}[t!]
\centering
\includegraphics[scale=0.61]{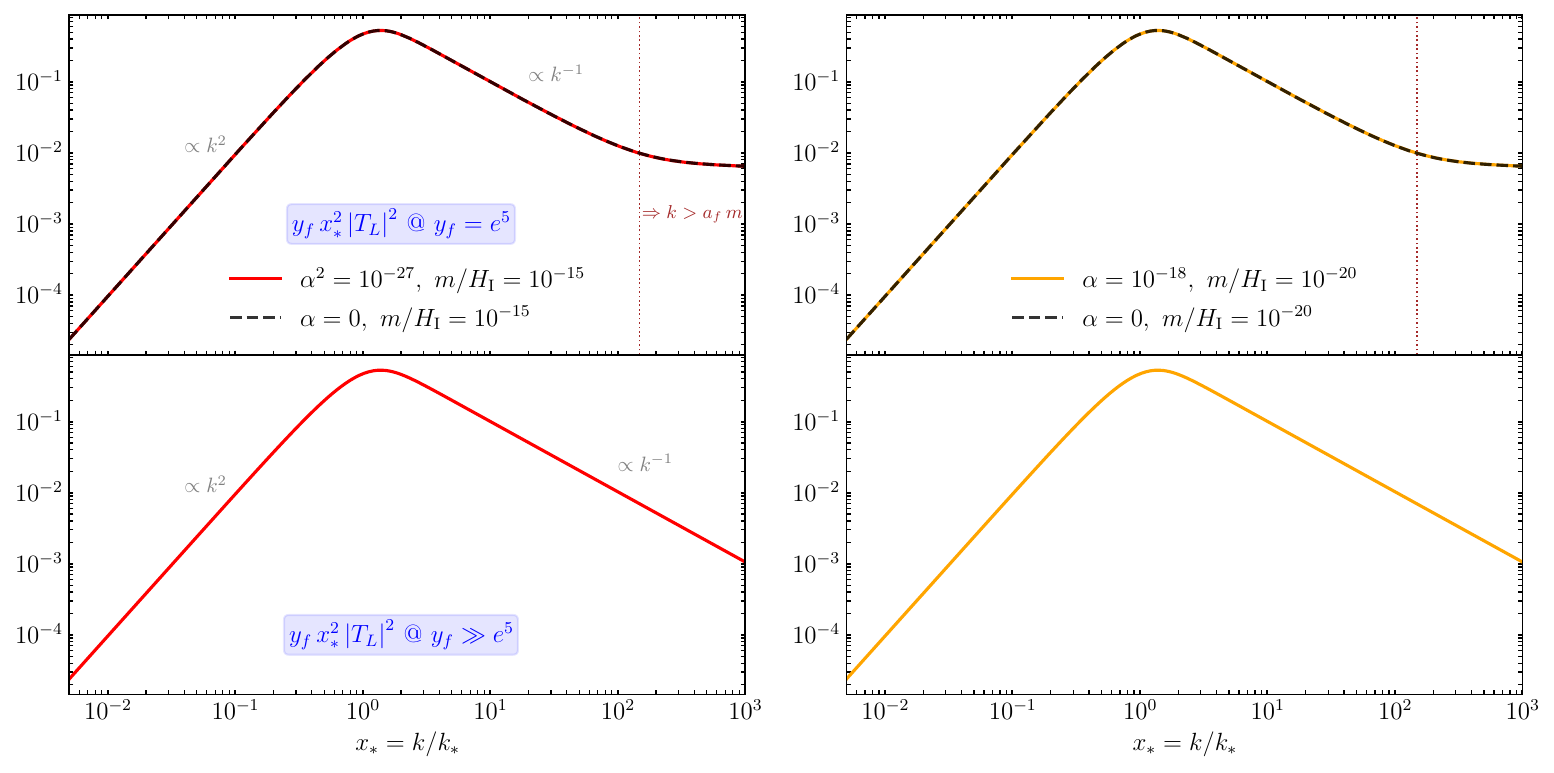}
\caption{Re-scaled power spectrum $y_f\, x_*^2\, |T_L|^2 $ (see \eg \eqref{LPS}) of the longitudinal modes evaluated for two different parameter choices (left and right panels) evaluated at 5 e-folds after $H(a_*) = m$, \ie when $\ln (y_f) = \ln(a_f/a_*) = \Delta N = 5$ (top) and much later times $y_f \gg {\rm e}^5$ (bottom). In the top panels, modes that reside on the right hand side of the dashed vertical line are still relativistic at the time $a_f$ of the evaluation satisfying $k > a_f\, m \to k/k_* = e^{\Delta N} =e^{5}$. At later times, the re-scaled power spectrum preserves its amplitude, however more modes shown in the UV tail becomes non-relativistic, adopting the $k^{-1}$ behavior of the modes around the peak.\label{fig:PSAL}}
\end{figure}
\begin{figure}[t!]
\centering
\includegraphics[scale=0.61]{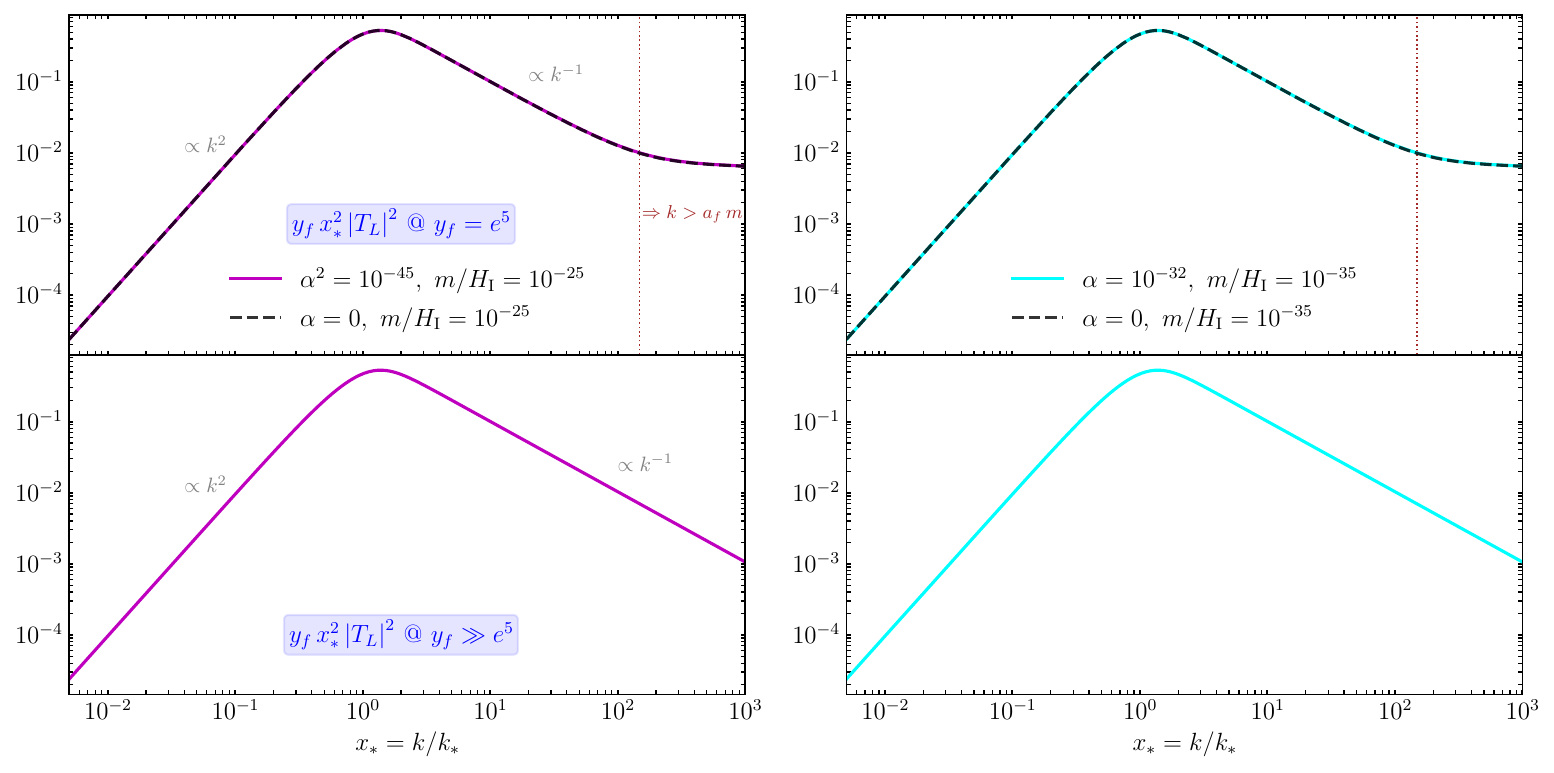}
\caption{Same as  Figure \ref{fig:PSAL}, but  with different parameter choices.\label{fig:PSAL2}}
\end{figure}
As the modes evolve from super-horizon to sub-horizon scales, we study the evolution of $T_L$ 
during RDU. We find convenient for our numerical
 implementation to introduce a new time variable $y$,
  and a quantity $x_*$ related with the wave-number $k$
 as
 \beq
 y\equiv\frac{a}{a_*}\hskip0.7cm,\hskip0.7cm x_* \equiv k \tau_* = \frac{k}{k_*}\,.
 \eeq
 In terms of these variables, we use the EoM \eqref{ALandAt} for the longitudinal modes to obtain the evolution equation for $T_L$ as
\begin{align}\label{Tfeom}
T_L'' (y,x_*) &+ \left[\fr{1 - ({\alpha^2 H^2}/{m^2})}{1 +({\alpha^2 H^2}/{m^2})}\right]\left(\frac{x_*^2}{x_*^2 + y^2[1+({\alpha^2 H^2}/{m^2})]}\right) \fr{2}{y}\,\, T'_L (y,x_*) \\ \nn
&\quad\quad\quad\quad\quad\quad\quad\quad\quad\quad +  \left[\fr{3 - ({\alpha^2 H^2}/{m^2})}{3 \left( 1 +({\alpha^2 H^2}/{m^2})\right)}\right]\left(x_*^2 + y^2 [1 + ({\alpha^2 H^2}/{m^2})]\right)T_L(y,x_*) = 0\,,
\end{align}
where $H$ is the Hubble rate during RDU \eqref{evhrd}: $H(y)= H_{\rm I} \left( y_{\rm end}/{y}\right)^2$ with 
\beq
\label{def_yend}
y_{\rm end}\,\equiv
\,\frac{a_{\rm end}}{a_*}\,=\,\sqrt{\frac{m}{H_I}}\,,
\eeq
 denoting the epoch signalling end of inflation, and the onset of RDU. Notice from the terms proportional to $\alpha$ in \eqref{Tfeom} that, the role of the non-minimal coupling with gravity can be important at early times during RDU due to the hierarchy $\alpha H_{\rm I}/m$, but its role 
 quickly becomes negligible
 as the universe expands and $y\gg y_{\rm end}$.

Using accurate initial conditions we derive on the inflation side (see Appendix \ref{AppB}), we numerically evolve \eqref{Tfeom} to obtain the power spectrum \eqref{LPS} evaluated at late times corresponding to $y_{f} = {\rm e}^{5}$ (\eg 5 e-folds after the epoch $H = m$) and $y_f \gg {\rm e}^{5}$. 
We show in  Figs. \ref{fig:PSAL} and \ref{fig:PSAL2} the resulting (re-scaled) power spectrum $y_f\,x_*\, |T_L|^2$
for
  representative parameter choices.  The power spectrum exhibits a peak amplitude at an intermediate momentum  $k_* = a_* m$ corresponding to the scale that re-enters the horizon when $H(a_*) = m$. The $k^2$
  rise of the spectrum from large towards small scales is inherited from the inflationary initial conditions (see section \ref{sec_ps1}). It is   due to the  constant behaviour of the transfer function $T_L$
  for long wavelength modes that mostly evolve at super-horizon scales. 
   On the other hand, smaller scale modes with wave-numbers $k \gtrsim k_*$ start to experience sub-horizon evolution with decaying amplitudes (as in region $({\bf H})$). At late times, this property causes  their power amplitude to decrease inversely proportional to the wave-number, $k^{-1}$. It is interesting to obtain with no fine-tunings a spectrum of fluctuations with a peak
   structure, a task that is not easy in the context of adiabatic fluctuations and primordial 
   black hole dark matter (see e.g. \cite{Ozsoy:2023ryl} for a recent review).

The profile for the longitudinal mode spectrum shown  in  Figs   \ref{fig:PSAL} and \ref{fig:PSAL2} shares the same qualitative behaviour of the Proca model (with $\alpha=0$) as  introduced  in \cite{Graham:2015rva}.
In fact, for not so large choices of $\alpha H_{\rm I}/m$ we adopt in Figures \ref{fig:PSAL} and \ref{fig:PSAL2}, the super-horizon dynamics of the vector modes around the peak $k_*$ is not much influenced  from  by the non-minimal coupling (see \eg Figs. \ref{fig:SAL} and \ref{fig:AL}), as they go through phases {\bf (A)-(E)-(F)} or {\bf (B)-(C)}.
    However, for larger values of the coupling $\alpha$ 
    we  increase the duration  of  phase {\bf (G)} (see eq. \eqref{DG}), hence the range of modes that enter the horizon during this phase.
    What is the fate of these modes? Can they contribute to the power spectrum in a dangerous way, especially for small scales that are not shown in Figure \ref{fig:PSAL}? What are the conditions we
    need to impose on the size of $\alpha$ in order to avoid catastrophic gradient
    instabilities?

\subsection{ The fate of the modes that re-enter the horizon during phase (G)}
\label{sec_phate}

We now discuss how to tame the instabilities
of short-scale modes going through
phase {\rm \bf (G)} of Fig \ref{fig:SAL}. (Recall the analytical discussion in section \ref{sec_grinan}.) These modes are dangerous because they might
considerably enhance the spectrum of vector modes at small
scales, generating a {\it second pronounced peak} besides 
the one we met in Fig \ref{fig:PSAL}. In this section we discuss
how to avoid this fact to occur. 

\smallskip

First of all, to understand the behavior of these modes, 
we  compute the range of vector wave-numbers affected by this epoch.  We evaluate the range of scales that re-enter the horizon between the end of inflation, $a_{\rm end}$, and the 
end of the dangerous phase  {\rm \bf (G)}, $a_c$. This range is $k_{\rm end}\le k \le k_c$. 
 In terms of the coordinate $y$, the  quantity $a_c$ is
\beq
y_c = \frac{a_c}{a_*} = \frac{a_c}{a_{\rm end}}\frac{a_{\rm end}}{a_*} \simeq \sqrt{\alpha},
\eeq
where we  use eqs  \eqref{DG} and  \eqref{def_yend}.
Therefore, the duration of the phase {\rm \bf (G)} is longer 
 for larger non-minimal coupling. In terms of the quantity $y_c$, the mode $k_c$ that re-enters the horizon at $a_c$ can be expressed as 
\beq\label{yc}
x^{c}_*\equiv \frac{k_c}{k_*} = \frac{a_c\, H_{\rm I}}{a_*\, m}\left(\frac{a_{\rm end}}{a_c}\right)^2 = y_c^{-1}\,.
\eeq
Hence, the range of wave-numbers that re-enters the horizon in phase {\rm \bf (G)} shifts towards larger and larger scales with increasing non-minimal coupling. Introducing the 
 wave-number 
 that leaves the horizon right at then end of inflation as
\beq
x^{\rm end}_* \equiv \frac{k_{\rm end}}{k_*} = \frac{a_{\rm end} H_{\rm I}}{a_* m} = y_{\rm end}^{-1} = \sqrt{\frac{H_{\rm I}}{m}},
\eeq 
we find that 
the range of wave-numbers that re-enter the horizon during the phase {\rm \bf (G)} can be summarized in terms of the model parameters as 
\beq
\label{instra1}
x^{c}_* < x_* < x^{\rm end}_*  \quad\quad \xRightarrow \quad\quad \frac{1}{\sqrt{\alpha}} < \frac{k}{k_*} < \sqrt{\frac{H_{\rm I}}{m}}.
\eeq
This expression indicates  that, choosing smaller and smaller values for $m/H_{\rm I}$, unstable modes will be pushed more towards UV, at a fixed non-minimal coupling. On the other hand at fixed vector field mass,  the unstable modes have a tendency to shift towards IR by increasing the non-minimal coupling with gravity.   

\begin{figure}[t!]
\centering
\includegraphics[scale=0.62]{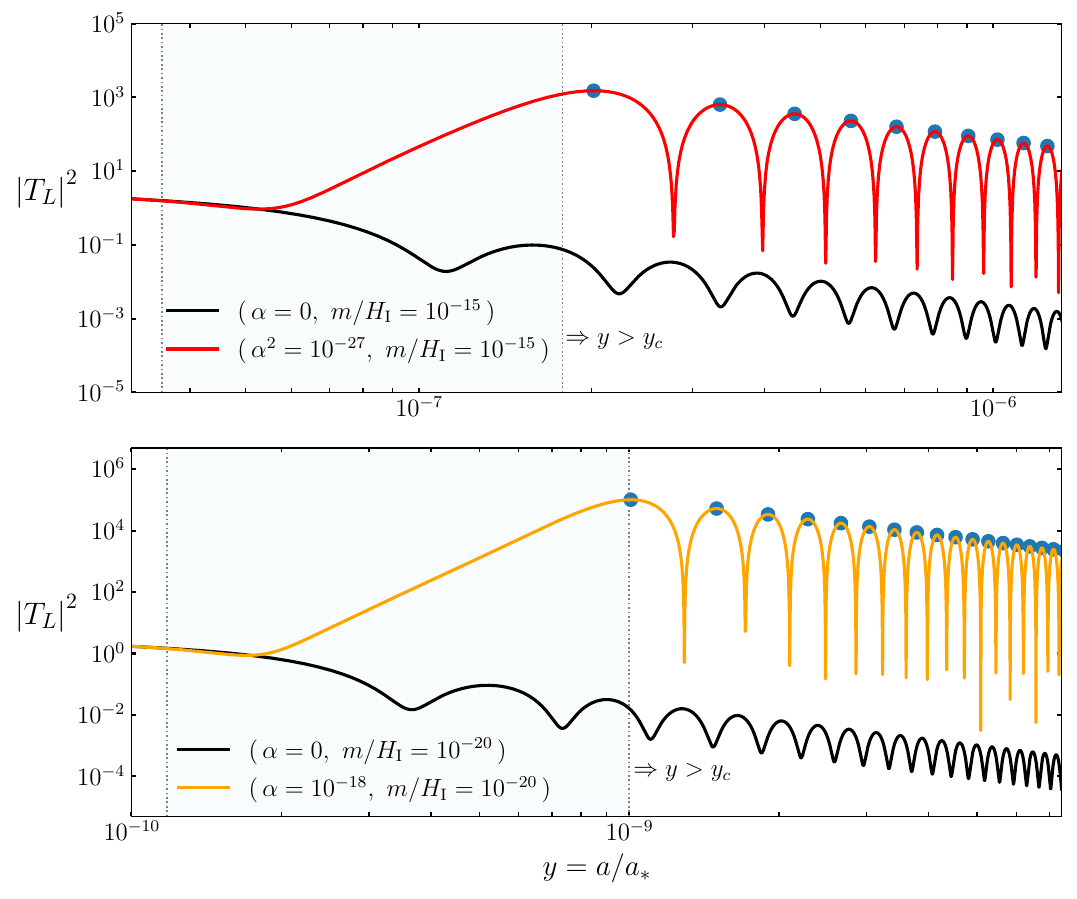}
\caption{Time evolution of the maximally enhanced longitudinal modes (with colors) that re-enters the horizon in phase {\rm \bf (G)}, corresponding to very small scales $x_* \simeq   2.8 \times 10^{7}$ (top-red) and $x_* \simeq  8.5 \times 10^{9}$ (bottom-orange) that exits the horizon just before the end of inflation, $ x_* \lesssim x^{\rm end}_*$. The part of the graphs highlighted by light blue indicates the time a mode spends in the unstable region as it enters inside the horizon. The scenario shown in the bottom panel has a larger $\alpha H_{\rm I}/m$ corresponding a longer {\bf (G)} phase (see eq. \eqref{DG}) such that a given short mode spends more time in the unstable region with an increasing amplitude. For $\alpha = 0$ (black curves), the unstable era is not present and these short scale modes starts to decay as soon as they re-enter the horizon (the left most vertical dotted line) \cite{Graham:2015rva}. \label{fig:SSALSq}}
\end{figure}
In Fig \ref{fig:SSALSq} we represent  two different scenarios to show the evolution of modes that leave the horizon right before the end of inflation.  These modes spend the most time within the unstable phase {\rm \bf (G)}: therefore the gradient instability maximally enhances their size  with respect to modes at larger scales.  The unstable short-scale modes enhance their size in proportion to the duration of phase {\rm \bf (G)}, as   determined by eq. \eqref{DG}. 

\smallskip
The danger -- as mentioned above --  is that the gradient instability acts on the small scale modes within the 
interval \eqref{instra1}, leading to a second peak in the spectrum more pronounced
than the first peak analyzed in Fig \ref{fig:PSAL}.   We wish to avoid this. 
Fortunately,  the mode  growth  stops as soon as they step inside  phase {\rm \bf (H)}, characterized by $y > y_c$. (See section \ref{sec_grinan}.) It is precisely this property that allows us to tame the instabilities.
At this stage, indeed,  they start to decay $|T_L| \propto a^{-2}$, spending a large amount of in the {\rm  \bf (H)} era, until they become non-relativistic at very late times: 
\beq\label{ynr}
 y_{\rm nr} = \frac{a_{\rm nr}}{a_*} = x_* = \frac{k}{k_*} \gg 1.
\eeq
The smaller the bare mass of the vector field is, the larger the amount of time these small scale modes spend inside the horizon in the decaying oscillatory phase {\rm \bf (H)}. This can be visually realised by comparing the overall scales in the horizontal  axis of Figure \ref{fig:SSALSq}, and by noticing  that for smaller bare mass,\ie $m/H_{\rm I}$, individual modes tend to become non-relativistic later, see \eg Fig \ref{fig:AL}. 

\medskip
Collecting the previous considerations, we can now estimate  whether  the
size of the 
 aforementioned
 second instability  peak  can
 be set under control.
The parameter choice should
ensure that the instability interval $a_{\rm end}\,<\,a\,<\,a_{c}$ remains small, and at the same time, we have  a prolonged  phase
$({\bf H })$ following the unstable phase $({\bf G})$.  We consider  the examples studied  in Figures \ref{fig:PSAL} and \ref{fig:PSAL2}. 
  We extrapolate the behaviour of the transfer function to late times by fitting a decaying envelope to the maxima of oscillations of $|T_L|^2$, as shown in Fig. \ref{fig:SSALSq}. In this way, we compute the re-scaled power spectrum $ y\, x_*^2 |T_L|^2$ 
    at $y_{\rm nr}$ (defined in eq. \eqref{ynr}) for each parameter choices shown in Figures \ref{fig:PSAL} and \ref{fig:PSAL2}. 
\begin{table}
\begin{center}
\begin{tabular}{| c | c | c |}
\hline
\hline
\cellcolor[gray]{0.9}&\cellcolor[gray]{0.9} $y_{\rm nr}\, x_*^2\, |T_L|^2  $ &\cellcolor[gray]{0.9} $y_{\rm nr} \equiv x_* $  \\
\hline
\cellcolor[gray]{0.9}$\alpha^2 = 10^{-27},\, m/H_{\rm I} = 10^{-15}$& $ 0.002 $ & $2.8 \times 10^{7}$ \\\hline
\cellcolor[gray]{0.9}$\alpha^2 = 10^{-36},\, m/H_{\rm I} = 10^{-20}$ & $0.001$ & $8.5 \times 10^{9}$ \\\hline
\cellcolor[gray]{0.9}$\alpha^2 = 10^{-45},\, m/H_{\rm I} = 10^{-25}$ & $0.004$ & $2.6 \times 10^{12}$     \\\hline
\cellcolor[gray]{0.9}$\alpha^2 = 10^{-64},\, m/H_{\rm I} = 10^{-35}$ & $0.001$ & $2.5 \times 10^{17}$ \\\hline
\hline
\end{tabular}
\caption{The peak amplitude of the re-scaled power spectrum for the maximally enhanced mode $x_*$ that re-enters the horizon during the gradient instability phase {\bf (G)} for different parameter choices parametrized by the $(\alpha^2, m/H_{\rm I})$ pairs\label{tab:SSpowa}. See main text for additional
explanations.}
\end{center}			
\end{table}
We present these results in Table \ref{tab:SSpowa}, which informs us that the second peak generated at  small scales by the gradient instability 
is well smaller than the first  peak at $k_*$. Similar considerations and numerical checks
can be done for  other pairs of parameters $(\alpha, m/H_I)$, so to circumscribe the region of parameters
 which avoid large effects from the gradient instability \footnote{A short Python notebook file that can be utilized for this purpose can be reached from \href{https://github.com/oozsoy/NM_Vector-Field_CosmoEval}{github}.}. We will return to this point in the next Section,
 when discussing the final abundance of longitudinal vector modes. 
 
\section{Abundance of longitudinal vector modes}
\label{sec_ab}

We now collect the previous results, and compute the final abundance of longitudinal vector modes. We show
that they can constitute the entirety of dark matter, for a broad range of vector masses.
The final dark matter abundance depends on two parameters only, $m$ and $\alpha$, making this scenario 
very economical.
\smallskip

First of all, we compute the longitudinal-mode contribution to the universe energy density. 
The presence of the non-minimal coupling coupling complicates the computation of the the energy density contained in the longitudinal modes. However, as we show in Appendix \ref{AppC}, the contributions due to the non-minimal coupling in the energy-momentum tensor (EMT) becomes negligible when all the modes associated with the peak in the power spectrum becomes non-relativistic.
 In  particular, the corrections induced by the non-minimal interactions appear at  order
  $\alpha^2 H^2\, \mathcal{O}(A'^2, A^2)$. Their size is then
    negligible  compared to terms of order $m^2\, \mathcal{O}(A'^2, A^2)$ for $a \gg a_c$, as can be realized by noticing that $\alpha^2 H^2$ is  comparable to $m^2$ at around $a\sim a_c$, see eq  \eqref{DG}. This situation can  also be interpreted as a consequence of the null energy condition, which ensures that Hubble rate is always decreasing. The  effects introduced by to non-minimal coupling are maximal  during inflation (where $H$ decreases very slowly), but subsequently their size diminishes as  the Hubble rate decays fast during the post-inflationary era. Therefore, at late times the energy density contained in the fluctuations of the longitudinal modes is \cite{Graham:2015rva}, 
\beq\label{rhoL}
\rho_{A_L} = \frac{1}{2a^4} \int \d \ln k\, \left\{\frac{a^2 m^2}{k^2 + a^2 m^2}\mathcal{P}_{A_L'}(\tau,k) + a^2 m^2 \mathcal{P}_{A_L}(\tau,k)\right\}.
\eeq
When all the modes we consider are inside the horizon, the fluctuations are virialized so that all quantities contribute equally
 to the energy density in \eqref{rhoL}. Parametrizing these contributions in terms of the power spectrum, and using the time variable $y = a/a_*$  adopted in the previous sections, we can therefore re-write $\rho_{A_L}$ in terms of the transfer function \eqref{Alrdu} of the longitudinal modes as 
\beq\label{rhoAL}
\rho_{A_L} = \frac{H_{\rm I}^2}{4\pi^2} \frac{a_* k_*^2}{m_{\rm eff,I}^2}\frac{m^2}{a^{3}(t)} \int \d \ln x_* \left\{y\, x_*^2\, |T_L(y,x_*)|^2\right\}.
\eeq
As can be inferred from the evolution of longitudinal modes deep in phase {\bf(D)} (see section \ref{NRAL}), the integrand in this expression is time-invariant, as far as we evaluate the power spectrum at epochs when all the modes $x_*$ are in the non-relativistic regime. Therefore, considering the peaked profile for the power spectrum (as in Fig. \ref{fig:PSAL}) in \eqref{rhoAL}, and
using 
\beq
\label{expks}
a_* \simeq \sqrt{\fr{H_{\rm eq}}{m}}\, a_{\rm eq}, \quad\quad k_{*} = a_* m \simeq \sqrt{H_{\rm eq}\, m}\,\, a_{\rm eq},
\eeq
the energy density in the longitudinal modes today gives
\beq
\rho_{A_L}(t_0) \simeq \, \frac{1.2}{4\pi^2} \frac{H_{\rm eq}^{3/2}}{(1+z_{\rm eq})^3}\,\frac{m^{5/2}}{\alpha^2},
\eeq
where we assume $m_{\rm eff,I}^2 \simeq \alpha^2 H_{\rm I}^2$ in the $\alpha H_{\rm I} \gg m$ regime we focus on. Taking into account the observed dark matter abundance $\rho_{\rm dm}(t_0) = 1.26 \times 10^{-6}\, {\rm GeV}/{\rm cm}^3$, we adopt the central value for the red-shift at equality $z_{\rm eq} = 3402$ \cite{Planck:2018vyg}. We  finally obtain the fraction of longitudinal modes in dark matter density today as
\beq\label{ALA}
f_{A_L}\equiv\frac{\Omega_{A_L}}{\Omega_{\rm dm}} \simeq \left(\fr{\alpha}{3.7 \times 10^{-22}}\right)^{-2} \left(\fr{m}{ 
1\, {\rm eV}} \right)^{5/2}.
\eeq

Importantly,
  the final longitudinal vector field abundance of eq. \eqref{ALA} is not sensitive to the energy scale $H_{\rm I}$ during inflation, nor to details of the underlying cosmology. Hence, the  dark matter abundance depends
  on two parameters only, the same that appeared in the initial vector action \eqref{sgf}.   The reason of this nice property is summarised in the discussion around eq. \eqref{infPSA}: thanks
  to contributions from non-minimal couplings to gravity, cancellations occur and the  amplitude
  of the inflationary spectrum of longitudinal modes is independent from $H_{\rm I}$.
 
\begin{figure}[t!]
\centering
\includegraphics[scale=0.66]{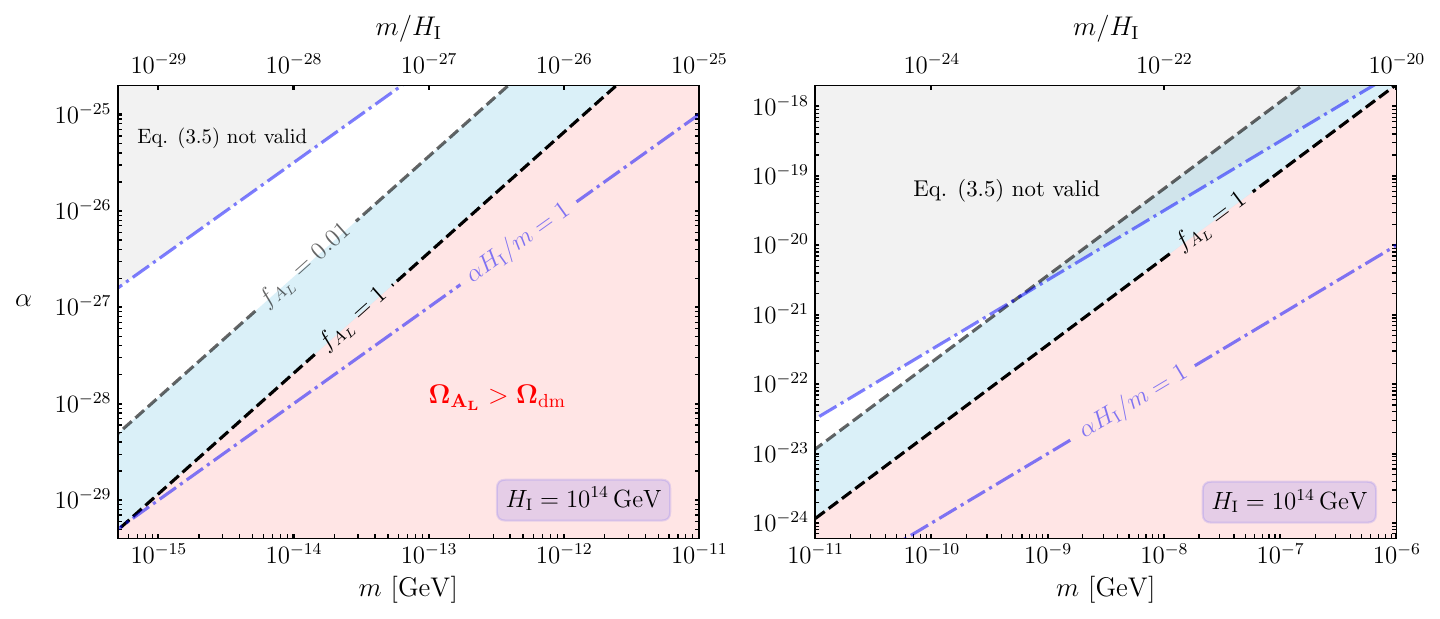}
\caption{Dark matter abundance from non-minimally coupled vector field in the $\alpha - m$ plane. The red region produces too much dark matter and should be excluded. The parameter space we are interested in lies above lower blue (dotted dashed) line where the non-minimal coupling is sizeable. The region highlighted by gray is also excluded on the grounds that eq. \eqref{ALA} is not valid within that parameter space where $\alpha H_{\rm I}/m > 10^5 - 10^4$ (see the discussion in the main text). To illustrate these limitations on the parameter space, we adopted a high scale inflationary model $H_{\rm I} = 10^{14} \,{\rm GeV}$ in both panels.\label{fig:fAL1}}
\end{figure}

\medskip

\begin{figure}[t!]
\centering
\includegraphics[scale=0.66]{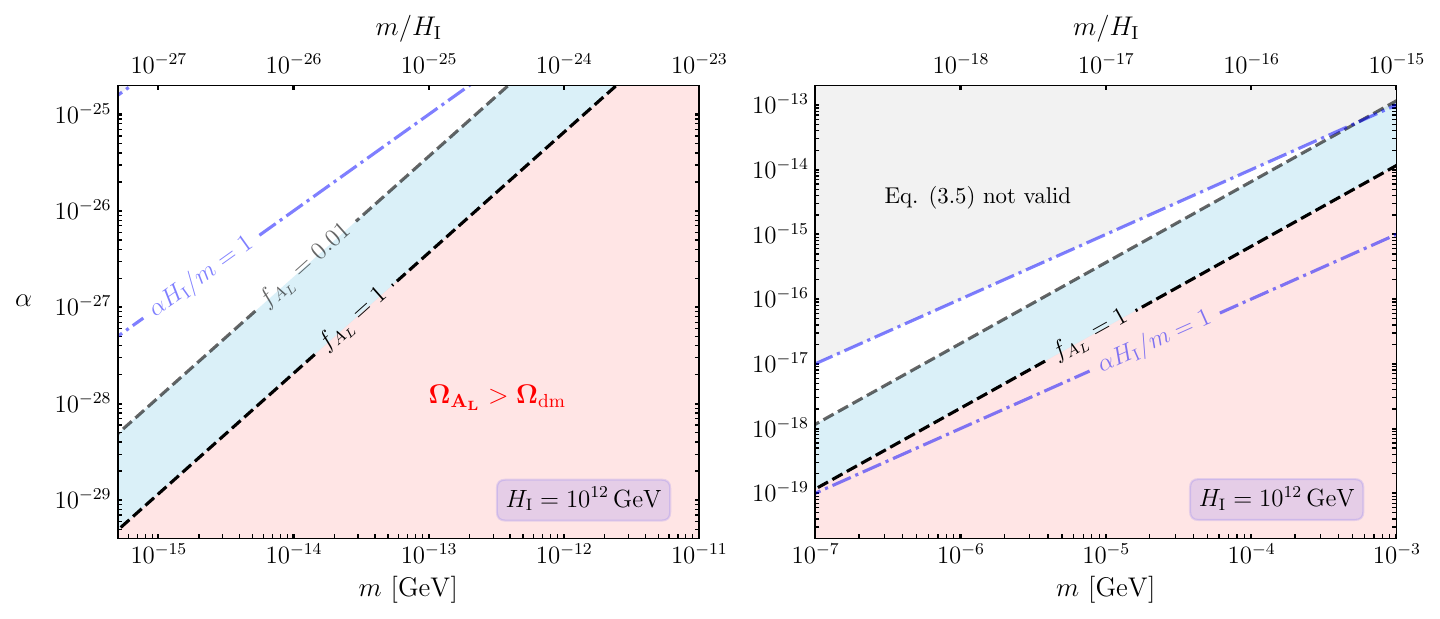}
\caption{Same as Figure \ref{fig:fAL1} for an inflationary scenario with $H_{\rm I} = 10^{12}\,{\rm GeV}$. In the right panel, the boundary of the excluded region (gray area) is parametrized by the $\alpha H_{\rm I}/m = 10^4$ for the shown mass range.\label{fig:fAL2}}
\end{figure}

Although there is no explicit dependence on $H_{\rm I}$, the final abundance implicitly depend on the scale of inflation, since  the validity of \eqref{ALA} assume a limiting value of $\alpha H_{\rm I}/m$ in order to tame the instability of the longitudinal modes in phase {\rm \bf (G)}, and to avoid a second large peak in the power spectrum of the latter at very short scales $k \geq k_{\rm c} \gg k_*$ (see the discussion in section \ref{sec_phate} and Table \ref{tab:SSpowa}).  Hence, to trust  our expression \eqref{ALA}, we can only increase the non-minimal coupling by a certain amount at fixed vector field mass $m$. For the range of $m/H_{\rm I}$ values we consider in Figures \ref{fig:fAL1} and \ref{fig:fAL2}, the limiting lines of $\alpha H_{\rm I}/m$ is shown by the upper dot dashed lines which rules out the gray shaded regions. Moreover, another limitation on the parameter space $(\alpha, m)$ stems from the fact we are interested in the sizeable non-minimal coupling regime parametrized by the inequality $\alpha H_{\rm I }/m > 1$, so to ensure the validity of eq. \eqref{infPSA}. We indicate this lower bound by the lower dot dashed lines labelled by $\alpha H_{\rm I}/m = 1$ in the same Figures. Therefore, the region between the dot dashed lines in Figures \ref{fig:fAL1} and \ref{fig:fAL2} indicates the region where \eqref{ALA} is applicable and the longitudinal modes can potentially constitute all of dark matter.
  
  To illustrate these constraints, we plot in Figure \ref{fig:fAL1}  the fractional abundance in the $(\alpha, m)$ parameter space by focusing on a high-scale inflationary scenario. We find that longitudinal modes can account for dark matter abundance for a large range of masses between 
  \begin{equation}
  \label{permran}
  5 \times 10^{-7} \lesssim m\, [{\rm eV}] \lesssim 5 \times 10^{3}\,,
  \end{equation} 
  after satisfying the aforementioned requirements. 
  This is in contrast with the Proca field studied in \cite{Graham:2015rva}, where the allowed vector field mass depends on the scale of inflation hence it is limited by constraints
  from 
   CMB observations. In fact,  the non-minimal coupling $\alpha$ we consider provides additional leverage on the parameter space making vector field a viable dark matter candidate for a  sizeable range of masses, including values with additional promising venues for detection.
   At the same time, our scenario is very economical since the dark matter
   abundance depends on two parameters only. 
     We stress
  that for deriving our conclusions we do not make any special hypothesis on the cosmological
  behaviour of the universe between inflation and radiation domination. We only assume
  a rapid and effective reheating process. 
  \smallskip
  
As can be realized from the left panel of Figure \ref{fig:fAL1}, it is not possible to lower the vector dark matter mass below $m \lesssim 5 \times 10^{-7}\, {\rm eV}$. This remains true if we lower the scale of inflation.  The region of interest $\alpha H_{\rm I}/m > 1$ tends to be realized for larger values of the non-minimal coupling $\alpha$, restricting $f_{A_L} = 1$ line to lie in the $\alpha H_{\rm I}/m < 1$ region of the parameter space where \eqref{ALA} does not apply. We illustrate these considerations in Figure \ref{fig:fAL2}. As shown in the right panel of the Figure, sufficient
dark matter abundance  can only be realized for relatively large vector masses. In general,  this trend becomes more and more apparent for scenarios that exhibit smaller inflationary scales. 

\medskip

To summarize this section, by adding non-minimal couplings
to gravity to the scenario of \cite{Graham:2015rva}, we determined a longitudinal vector dark matter
model
 where the dark matter abundance depends on two parameters
only, $\alpha$ and $m$, see eq. \eqref{ALA}  (We
show in Appendix \ref{sec_trab} that the amplitude of transverse modes remain subdominant).  This scenario is very economical; nevertheless, the expression
for the dark matter
abundance \eqref{ALA} leaves  some degeneracy because the parameters $\alpha$
and $m$ can be chosen arbitrarily, at least within the mass interval \eqref{permran}.
In the next section we discuss further phenomenological consequences of our framework,
aimed to distinguish the effects of the two parameters $(\alpha, m)$ on observable
quantities,  which
might allow us to obtain independent measurements of their size.

\section{Phenomenological considerations}
\label{sec_pheno}

We learned in the previous section that our version of the scenario introduced
in \cite{Graham:2015rva} realizes dark
matter in the form of longitudinal modes of a massive vector non-minimally coupled
with gravity. The dark matter abundance depends on two parameters only, making
the model  very economical. In order to provide the totality of dark matter, the non-minimal
coupling $\alpha$ -- as appearing in action \eqref{sgf} -- and the vector mass $m$ are related by (see eq. \eqref{ALA}) 
\begin{equation}
\alpha \simeq  3.7 \times 10^{-22}\,\left( \frac{m}{1\,{\rm eV}} \right)^{5/4}\,.
\end{equation}
Hence, for relatively light vector fields (recall the allowed mass range of eq.\eqref{permran}), 
the non-minimal coupling  
 of vector to gravity of
 in action \eqref{sgf}  acquires  very small values \footnote{It would
 be interesting to investigate whether
 such small values of $\alpha$
 are stable under loop corrections. In fact,
 when  working in a decoupling limit,
  the  vector couplings 
 we consider are close relatives to Galileons (see \cite{Tasinato:2014eka,Heisenberg:2014rta}).
 The latter 
 are known to satisfy non-renormalization  theorems \cite{Luty:2003vm,Nicolis:2004qq}, which
 can also be extended to curved space \cite{Pirtskhalava:2015nla}.}. At the same
time, the vector mass (for high-scale inflation) can be chosen within the interval
\eqref{permran}, leaving some degeneracy in parameter space for realizing scenarios
where the totality of dark matter is constituted by longitudinal modes.

\medskip
Besides 
 cosmological
consequences of dark photon dark matter,
we now ask whether there are any further experimental avenues to independently probe the parameter
$m$  appearing in action  \eqref{sgf}. Let us consider  scenarios whose
tree level actions do not contain 
 gauge interactions between  visible and   dark sectors, besides
gravity. Nevertheless, loop effects inevitably
induce  portal couplings with the Standard Model, 
for example through a kinetic mixing with the SM photon $A_{(\gamma)\,\mu}$ \cite{Holdom:1985ag,Fayet:1980ad,Fayet:1980rr,Dobrescu:2004wz}
\beq
\label{exp_kinm}
{\cal L}_{\rm km}\,=\,-\frac{\varepsilon}{2} F_{\mu\nu}F^{\mu\nu}_{(\gamma)}\,.
\eeq
 $F^{\mu\nu}_{(\gamma)}$ is the field strength for the SM photon field, while $F_{\mu\nu}$
the dark vector field strength. This gauge-invariant dimension-four operator is controlled by the dimensionless
parameter $\varepsilon$, which induces
 milli-charged couplings of dark photons to ordinary matters.
 Constraints exist on $\varepsilon$, which depend
on the dark photon mass (see e.g. \cite{Fabbrichesi:2020wbt} for a comprehensive review). 
Even for set-up where there are 
{\it no} tree-level gauge couplings between visible and  dark sectors, 
it has been demonstrated (see \cite{Gherghetta:2019coi}) that 
 loop effects mediated by
gravity alone
generate kinetic mixing of the form \eqref{exp_kinm}. The resulting values of the parameter
 $\varepsilon$ are very tiny: $\varepsilon\simeq 10^{-13}$; nevertheless
 such small values can be probed for dark vector masses in the ${\rm eV}-{\rm keV}$ range \cite{An:2014twa} (see also \cite{Pospelov:2008jk,Bloch:2016sjj}),  which is within the mass interval \eqref{permran} we found. 
  Hence, direct experiments testing
  milli-charged couplings with the dark sector can independently probe the
  range of vector masses  which can be realized through non-minimal
  couplings of the dark vector to gravity.

\medskip
Another option is to make use of  gravitational wave  experiments. As mentioned in 
the introduction, it is conceivable  to use gravitational wave  detectors to directly probe  effects of space-time deformations
associated with vector degrees of freedom. See e.g. \cite{Nomura:2019cvc} who
 uses a vectorial version of the so-called
 Khmelnitsky-Rubakov effect \cite{Khmelnitsky:2013lxt} in the context
 of pulsar timing-array experiments.
 
  Alternatively, one can
study primordial black holes  \cite{Passaglia:2021jla,Yoo:2021fxs} 
or induced  stochastic
gravitational wave   backgrounds (SGWB) \cite{Domenech:2021and}
produced at second order in perturbations by large-size  iso-curvature modes~\footnote{Recall that the
longitudinal vector  degrees of freedom correspond to iso-curvature fields, see section \ref{sec_ps1}.}. This is particularly interesting for vector mass ranges in the ${\rm eV}-{\rm keV}$ interval,
as we are going to discuss. The subject of scalar-induced SGWB
by adiabatic fluctuations, first investigated in 
\cite{Matarrese:1993zf,Matarrese:1997ay,Nakamura:2004rm,Ananda:2006af,Baumann:2007zm,Saito:2008jc,Saito:2009jt,Kohri:2018awv,Espinosa:2018eve,Inomata:2019yww},
is very well developed by now (see the review \cite{Domenech:2021ztg}). Much less studied are SGWB
  induced by iso-curvature modes -- the case relevant
for us since we study  vector longitudinal modes corresponding
to iso-curvature fluctuations. A 
 notable exception is \cite{Domenech:2021and}, where  the authors carry on a detailed
 analysis of how an enhanced spectrum of iso-curvature modes at super-horizon
 scales can induce a stochastic GW background after re-entering the horizon in
a  radiation-dominated universe. 
 This is possible because the early iso-curvature modes can get converted into curvature fluctuations upon horizon crossing, which in turn
 source  GWs. 
Therefore, similar to the well studied adiabatic case an enhanced, peaked spectrum
of longitudinal iso-curvature modes (with a maximum at scale $k_*$) could lead to a stochastic GW background peaked at scales of the same order of magnitude. Since in our case the scale of the scalar  peak is proportional to the vector mass $k_*=a_* m$ (see eq. \eqref{expks}), converting this quantity to the frequency domain ($f=k/(2\pi)$), we expect to generate a SGWB enhanced
at a peak frequency   
\begin{align}
f_* & \simeq   10^{-3} \left(\frac{m}{\rm   eV}\right)^{1/2}\,{\rm Hz}.
\label{lisa_fr}
\end{align}
A choice of
vector dark matter mass around the ${\rm eV}-{\rm keV}$ range -- as can be realized in our scenario \eqref{permran} -- can then enhance the SGWB signal in the milli-Hertz band, making it potentially detectable
by the LISA interferometer \cite{LISACosmologyWorkingGroup:2022jok}. 

 \begin{figure}
\centering
\includegraphics[scale=0.65]{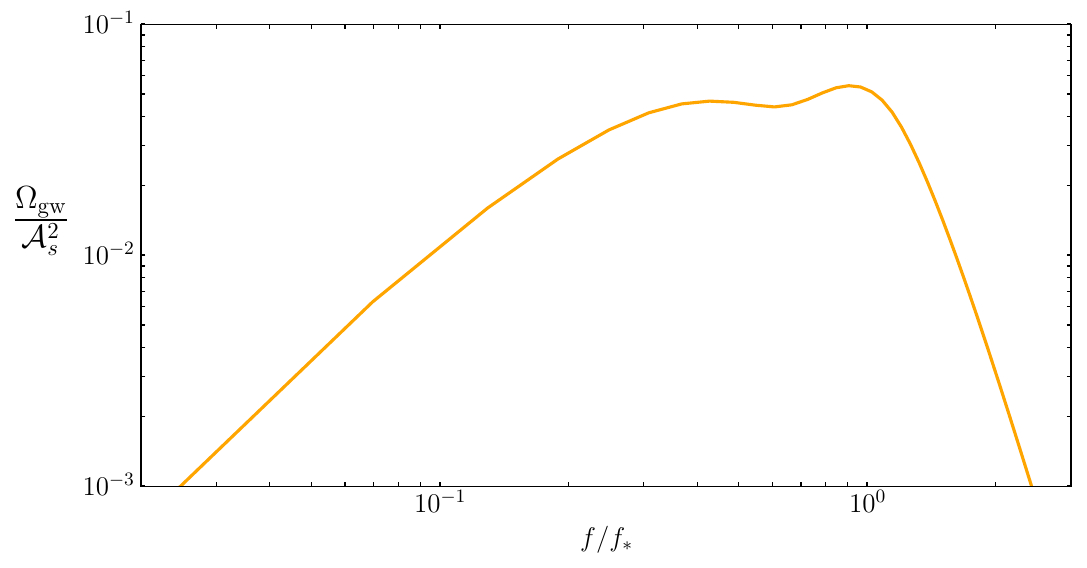}
\caption{The GW density spectrum induced at second order
by longitudinal vector iso-curvature fluctuations. See the main text, in particular the discussion after eq. 
 \eqref{lisa_fr} for a description of the set-up.} 
\label{fig:OGW}
\end{figure}

We realize that
the formulas developed in \cite{Domenech:2021and}   {\it can not} be directly
used in our context, without  some extra work. The derivations of  \cite{Domenech:2021and}
take into due account the back-reaction of iso-curvature fluctuations to gravity, 
while in this work we do not consider the back-reaction of vector
fields in the geometry. Moreover, the formulas of  \cite{Domenech:2021and}
assume  that the  
primordial iso-curvature spectrum is evaluated in a super-horizon regime, while
the  peaked shape of the spectrum presented
in Figs \ref{fig:PSAL} and \ref{fig:PSAL2} include also contributions from modes
evolving at sub-horizon scales.  
Nevertheless, we expect that a future, more careful
analysis of energy fluctuations associated
with iso-curvature modes (including back-reaction
in the geometry) again will lead
to a spectrum exhibiting a peaked structure, with
a peak around the scale $k_* = a_* m$
found in the previous section. The analysis
of iso-curvature density fluctuations
carried on in \cite{Graham:2015rva} indicates that
the corresponding spectrum grows as 
$\left( k/k_* \right)^3$
from large towards small scales, up
to a maximum at around $k\simeq k_*$, to then
decrease as
$\left( k/k_* \right)^{-1}$ for smaller scales $k\ge k_*$. 
We take such a spectrum profile as representative Ansatz 
of longitudinal vector density fluctuations, and denote ${\cal A}_s$ as the
maximum of the spectrum at $k = k_*$. We then numerically compute the
integrals of \cite{Domenech:2021and}, using their kernels and obtain
the profile for the GW density $\Omega_{\rm gw}$ as a function of $f/f_*$ as
represented in Fig \ref{fig:OGW}. The spectrum has a peaked structure, with a relatively
broad maximum at scales corresponding to the frequency region around the $f_*$  of eq. \eqref{lisa_fr} detectable with LISA. In the IR scales shown, it has a relatively mild slope as compared to GWs induced by adiabatic fluctuations ($ \Omega_{\rm gw} \sim f^3$) with a $f^{<2}$ behavior, while in the UV it decays as fast as $f^{-6}$.
It would be interesting to perform a more careful analysis on the shape and amplitude ${\cal A}_s$ in a physically realistic set-up as it could potentially provide us information about the couplings and parameters in our proposal \eqref{sgf}. We leave this important task to future studies. 

\section{Conclusions and Outlook}
\label{sec_dis}

We proposed a scenario of dark matter in the
form of longitudinal massive vector bosons,
 extending the  work \cite{Graham:2015rva} by including
 non-minimal couplings of the vector
 with gravity. Our set-up, besides
 being well motivated from an effective field theory perspective, is very economical
 since the final dark matter abundance depends
 on two parameters only (see eq. \eqref{ALA}): the vector
 mass and its non-minimal coupling with gravity. 
We determined and analysed in detail the rich dynamics of vector fluctuations
during inflation and radiation domination. In particular, we showed that during inflation, quantum fluctuations of longitudinal modes obtain a  large super-horizon power, which depends on the inverse strength of the non-minimal coupling with gravity. The latter induces gradient instabilities on vector field dynamics during radiation dominated era, which can nevertheless be tamed. For weak non-minimal coupling with gravity, the inflationary era thus sets suitable initial conditions for the longitudinal component of vector field. The resulting post-inflationary vector energy density can  account for the totality of dark matter, for a wide range of vector masses \eqref{permran}. We emphasize that the set-up we described is also minimal in the sense that it assumes standard post-inflationary history following the inflationary epoch. It would also
be interesting to explore variations of our proposal on dark matter production by utilizing non-trivial post-inflationary histories beyond the instantaneous reheating approximation we undertake in this work, as done in \cite{Kolb:2017jvz, Ahmed:2020fhc}.

Due to non-minimal coupling, the resulting abundance of dark matter leads to a wider range of allowed vector masses with respect to \cite{Graham:2015rva}. Interestingly, such mass range can be experimentally probed
by means of induced gravitational waves in the LISA frequency band. We  touched this argument in section \ref{sec_pheno} in a phenomenological way: it certainly deserves further investigations.

Given the absence of direct evidence of dark matter besides its gravitational
interactions with visible matter, it is perhaps a good idea to focus on gravitational experiments as dark matter detectors.
Gravitational wave experiments can provide new ways to probe light and feebly interacting dark matter (see e.g. \cite{Bernal:2017kxu} for a review). It will important in the near future to further explore the opportunities they can provide.

\acknowledgments
It is a pleasure to thank Michele Cicoli, Matteo Licheri, and Francisco Pedro for useful discussions. O\"O would like to thank Perimeter Institute for Theoretical Physics for hospitality while some part of this work is completed. O\"O is supported by the “Juan de la Cierva” fellowship IJC2020-045803-I and by the Spanish Research Agency (Agencia Estatal de Investigación) 
 through the Grant IFT Centro de Excelencia Severo Ochoa No CEX2020-001007-S, funded by MCIN/AEI/10.13039/501100011033. GT  is partially funded by the STFC grants ST/T000813/1
and ST/X000648/1. For the purpose of open access, the authors have applied a Creative Commons Attribution licence to any Author Accepted Manuscript version arising.

\begin{appendix}
\section{Equations of motion and mode functions of the gauge field}\label{AppA}
In this appendix, we derive the equations of motion of the gauge fields following two different methods: i) first by expanding the action \eqref{sgf} and decomposing the vector field into its temporal and spatial components ii) by following a covariant approach we will discuss below. We begin with the former by expanding the kinetic term of the gauge fields as 
\begin{align}\label{FSK}
\nn \fr{1}{4}F_{\mu\nu}F^{\mu\nu} &= \fr{1}{2}\,g^{00}\,g^{ij}\,F_{0i}F_{0j} + \fr{1}{4} \,g^{ij}\,g^{kl} F_{ik}\, F_{jl},\\
& = -\fr{1}{2a^2} (\dot{A}_i - \partial_i A_0)^2 + \fr{1}{4a^4}\, \delta^{ij}\,\delta^{kl} \epsilon_{mik} (\vec{\nabla}\times \vec{A})_m \, \epsilon_{njl} (\vec{\nabla}\times \vec{A})_n,
\end{align}
where we decomposed the non-vanishing component of the field strength tensor as  $F_{0i} = - F_{i0} = \dot{A}_i - \partial_i A_0$ and $F_{ij} = \epsilon_{lij} (\vec{\nabla}\times \vec{A})_l $. Using \eqref{FSK} in \eqref{sgf}, we re-write the action \eqref{sgf} as
\beq\label{sgfa}
S_{\rm GF} = \fr{1}{2}\int \d^4 x\, a^3 \left\{\fr{(\dot{A}_i - \partial_i A_0)^2}{a^2} - \fr{(\vec{\nabla}\times \vec{A})^2}{a^4} + m^2 \left(A_0^2 - \fr{A_i A_i}{a^2}\right) + \fr{\alpha^2}{3} \left(G^{00} A_0^2 + G^{ij} A_i A_j\right)\right\}.
\eeq
Notice that the action does not contain time derivatives of the temporal component of the gauge field which informs us that $A_0$ is a Lagrange multiplier whose equation of motion indicates a constraint of the system.  
To derive the latter, we require the knowledge of the Einstein tensor at the background level. Using the Einstein equation by adopting an energy momentum tensor of a barotropic perfect fluid with a constant equation of state $P = w \rho$, we have
\beq\label{EE}
G_{\mu\nu} \equiv R_{\mu\nu} - \fr{1}{2}R g_{\mu\nu} =  \fr{T_{\mu\nu}}{\Mp^2} = (1 + w)\fr{\rho}{\Mp^2}\,\, u_{\mu} u_{\nu} + w\fr{\rho}{\Mp^2}\, g_{\mu \nu}
\eeq
where $u_\mu = (-1, \vec{0})$ is the light-like four velocity of the fluid satisfying $u^{\mu} u_{\mu} = -1$. Finally, utilizing the Friedmann equation $3H^2\Mp^2 = \rho$ in \eqref{EE}, we found 
\beq\label{ET}
G_{\mu\nu} = 3H^2 \bigg[(1 + w)\,u_{\mu}u_{\nu} + w \,g_{\mu\nu}\bigg].
\eeq
We then plug \eqref{ET} in \eqref{sgfa} and vary the resulting action with respect to $A_0$ to obtain the following formal equation for the constraint as
\beq\label{con}
A_0 = - \left[a^{2} (m^2 + \alpha^2 H^2) - \vec{\nabla}^2\right]^{-1} \partial_i \dot{A}_i.
\eeq
Re-inserting the constraint back in \eqref{sgfa}, we re-write the action in terms of the $A_i$ that contains the physical degrees of freedom  as
\begin{align}\label{sgff}
S_{\rm GF} = \fr{1}{2}\int \d^3 x\, \d \tau \bigg\{A'_i A'_i - \partial_i A'_i \left[a^{2} (m^2 + \alpha^2 H^2) - \vec{\nabla}^2\right]^{-1} \partial_i A'_i &- (\vec{\nabla}\times \vec{A})^2 \\ \nn
&\quad - a^2(m^2 - w\,\alpha^2 H^2 )\, A_i \, A_i\bigg\}.
\end{align}
To disentangle the dynamics of the physical longitudinal and transverse modes of the vector field, we go to the Fourier space and decompose the spatial vector field into its different polarization states as
\beq\label{decomp}
A_i(\tau, \vbf{x})=\int\, \frac{\mathrm{d}^3 k}{(2 \pi)^{3 / 2}}\, \mathrm{e}^{i \vbf{k} \cdot \vbf{x}} A_i(\tau,\vbf{k}), \quad A_i(\tau,\vbf{k})\equiv \sum_{\lambda = L,\pm} \epsilon_i^{(\lambda)}(\vbf{k}) {A}_\lambda(\tau, \vbf{k})
\eeq
where noting the unit vector along $k_i$ as $\hat{k}_i$, the polarization vectors obey
\begin{align}\label{pve}
\nn &\hat{k}_i\, \epsilon_i^{(\pm)}(\vbf{k}) =0, \quad \epsilon_{i j k}\, k_j \,\epsilon_k^{(\pm)}(\vbf{k})=\mp i|\vbf{k}| \, \epsilon_i^{(\pm)}(\vbf{k}),\\ \nn & \hat{k}_{i}\, \epsilon^{(L)}_{i}(\vbf{k}) = 1,\quad \epsilon_{i j k}\, k_j \,\epsilon_k^{(L)}(\vbf{k})=0,\quad\quad\epsilon_i^{(\lambda)}(\vbf{k})\, \epsilon_i^{(\lambda^{\prime})}(\vbf{k})^* =\delta^{\lambda \lambda^{\prime}}, \\ 
&\epsilon_i^{(\pm)}(\vbf{k})^*=\epsilon_i^{(\pm)}(-\vbf{k})=\epsilon_i^{(\mp)}(\vbf{k}), \quad \epsilon_i^{(L)}(\vbf{k})^*=\epsilon_i^{(L)}(-\vbf{k})=\epsilon_i^{(L)}(\vbf{k}),
\end{align}
so that
\beq
\vbf{k}\cdot \vec{A}(\tau,\vbf{k}) = |\vbf{k}|\, A_L(\tau,\vbf{k}) = k\, A_L(\tau,\vbf{k}).
\eeq
Note that the relations in the last line of \eqref{pve} imply ${A}_\lambda(\tau, \vbf{k})^{*}= {A}_\lambda(\tau,-\vbf{k})$ for $\lambda = \{L,\pm\}$ so that ${A}_i(\tau, \vbf{x})$ is a real. Inserting the decomposition \eqref{decomp} into the action \eqref{sgff} and using \eqref{pve}, the quadratic action for mode functions of the transverse and longitudinal mode decouples as
\begin{align}\label{ASandT}
\nn S^{(2)}_L &= \frac{1}{2}\int \d^3 k\,\d \tau \left\{\frac{a^2 (m^2 + \alpha^2 H^2)}{k^2 + a^2 (m^2 +\alpha^2 H^2)} \big|A'_L (\tau,\vbf{k})\big|^2 - a^2 (m^2 - w\, \alpha^2 H^2) \big|A_L (\tau,\vbf{k})\big|^2 \right\} ,\\
S^{(2)}_T &= \frac{1}{2}\int \d^3 k\,\d \tau\,\sum_{\lambda = \pm} \left\{ \big|A'_{\lambda} (\tau,\vbf{k})\big|^2 -\left(k^2 + a^2 (m^2 - w\, \alpha^2 H^2)\right) \big|A_{\lambda} (\tau,\vbf{k})\big|^2 \right\},
\end{align}
Varying the actions \eqref{ASandT} above leads to the equations of motion (EoMs) \eqref{ALandAt} we introduced in terms of the conformal time.

\medskip
\noindent{\bf Covariant approach}
\smallskip

\noindent We can also follow a covariant approach to directly derive the equations of motion varying the action \eqref{sgf} with respect to $A^{\nu}$. In the following, we will re-write the action \eqref{sgf} in a form suitable to derive the equations of motion (EoM) in the configuration space. For this purpose, we first note that the kinetic terms $\propto F_{\mu\nu}F^{\mu\nu}$ in the action \eqref{sgf} can be re-written as
\beq\label{sgfalt}
S_{\rm GF}= \frac{1}{2}\int \d^4 x { \sqrt{-g}} \,\bigg[ -{(\nabla_\mu A_\nu)(\nabla^{\mu} A^{\nu})- A^\nu (\nabla_\mu \nabla_{\nu} A^{\mu}) - m^2 g_{\mu \nu} \,A^{\mu}A^{\nu}}+\frac{\alpha^2}{3}\, G_{\mu\nu}\, A^{\mu} A^{\nu}\bigg],
\eeq
where we utilized
\beq\label{Fsq}
-\frac{1}{4} F^{\mu\nu} F_{\mu\nu} \equiv -\frac{1}{2} (\nabla_\mu A_\nu)( \nabla^{\mu} A^{\nu}) + \fr{1}{2}({\nabla_\mu}{A_\nu})(\nabla^{\nu} A_\mu),
\eeq
by integrating by parts the second term in \eqref{Fsq}. Noting the definition of the Riemann tensor in terms of the commutator of the covariant derivatives
\beq\label{Rdef}
[\nabla_\mu, \nabla_\nu]\, A^{\rho} = R^{\rho}_{\,\, \sigma \mu \nu}\, A^{\sigma},
\eeq
and the Ricci tensor
\beq\label{Riccidef}
R^\rho_{\,\,\mu \rho \nu} \equiv R_{\mu\nu},
\eeq
we can then describe the second term in \eqref{sgfalt} in terms of the latter as 
\beq\label{id}
- A^\nu (\nabla_\mu \nabla_{\nu} A^{\mu}) \equiv - A^\nu \left(\nabla_\nu \nabla_{\mu} + R_{\mu \nu}\right) A^{\mu}.
\eeq
Plugging \eqref{id} back to the action \eqref{sgfalt} and performing a final integration by parts on the first term in the right hand side of \eqref{id}, we obtain
\beq\label{sgfff}
S_{\rm GF}= \frac{1}{2}\int \d^4 x { \sqrt{-g}} \,\bigg[ -(\nabla_\mu A_\nu)(\nabla^{\mu} A^{\nu})+ (\nabla_{\nu} A^{\nu})^2 -m^2_{\mu\nu}\,A^{\mu}A^{\nu}\bigg]
\eeq
where we defined the symmetric mass tensor of the vector field as 
\beq\label{masst}
m^2_{\mu \nu} \equiv m^2 g_{\mu \nu} + R_{\mu \nu} - \frac{\alpha^2}{3}\, G_{\mu\nu},
\eeq 
where the Einstein tensor given by eq. \eqref{ET}. 
Varying the action \eqref{sgff}, the EoM of $A_\nu$ can be obtained as 
\beq\label{EoM0}
\Box A_\nu - \nabla_\nu (\nabla^{\mu} A_\mu) - m_{\mu\nu}^2 A^{\mu} = 0,
\eeq
where $\Box \equiv \nabla^{\mu} \nabla_{\mu}$ is the d'Alembert operator. 
In summary, the equations of motion describing the dynamics of the vector field components can be derived using \eqref{masst}, \eqref{EoM0} by noting \eqref{ET} together with the following expressions of Ricci tensor and scalar for a given FRW background parametrized by a constant equation of state $w$ and Hubble rate $H = \dot{a}/a$:
\beq\label{Ricci}
R_{\mu\nu} = 3H^2\,\bigg[(1+w)\,\, u_{\mu} u_\nu + \fr{(1-w)}{2}\, g_{\mu\nu}\bigg] \quad \to \quad R = 3H^2 (1-3w).
\eeq

\medskip
\noindent{\bf Equations of motion}
\smallskip

\noindent To derive the equations satisfied by the vector field components, we work with cosmic time and the background metric ${g}_{\mu\nu} = {\rm diag} (-1, \vec{a}(\tau)^2)$. Using the standard definitions of the covariant derivative, the $\nu = 0$ component of the EoM \eqref{EoM0} can be found to give the constraint equation in \eqref{con}. On the other hand, the spatial components $\nu = j$ of the EoM \eqref{EoM0} gives
\begin{align}
\label{BAK} \ddot{A}_k + H \dot{A}_k-\bigg[\frac{\vec{\nabla}^2}{a^2}- \big(m^{2}-w\,\alpha^2{H}^2\big)\bigg] A_k - \partial_k \dot{A}_0 +  \frac{\partial_k(\partial_i A_i)}{a^2} - H \partial_k A_0 = 0.
\end{align}

\medskip
\noindent{\bf Helicity decomposition and mode functions}
\smallskip

\noindent Equation \eqref{BAK} contains the information about the dynamics of both the longitudinal and the transverse mode. The decouple their dynamics we decompose the components of the full vector field as 
\beq\label{b4}
A_\mu (\tau, \vbf{x}) = \int\fr{ \d^3 k}{(2\pi)^{3/2}}\, e^{i \vbf{k}.\vbf{x}} \sum_{\lambda = L, \pm} A^{(\lambda)}_{\mu} (t, \vbf{k}),\quad\quad A^{(\lambda)}_\mu (t,\vbf{k})^{*} = A^{(\lambda)}_\mu (t,-\vbf{k}),
\eeq
by defining the following expressions
\begin{align}\label{b6}
\nn A^{(L)}_{0} &= A_0(t,\vbf{k}), \quad A^{(L)}_i = \epsilon^{(L)}_{i}(\vbf{k}) \,A_L(t,\vbf{k}) ,\\
A^{(\pm)}_{0} &= 0 , \quad A^{(\pm)}_{i} = \epsilon^{(\pm)}_i(\vbf{k})\, A_{\pm} (t,\vbf{k}),
\end{align}
where the polarization vectors to obey the same relations \eqref{pve} as before. With these identifications, the Fourier space expression for the constraint equation \eqref{con} can be derived as  
\beq\label{conf}
A_0(t,\vbf{k}) = - i\fr{\, ({k}_i\,\epsilon^{(L)}_i(\vbf{k}))}{k^2 + a^2 m_{\rm eff}^2}\,\dot{A}_L(t,\vbf{k}),\quad\quad m_{\rm eff}^2 = m^2 + \alpha^2 H^2.
\eeq
Utilizing the decomposition (eqs. \eqref{b4} and \eqref{b6}) we defined above and the constraint \eqref{conf} in \eqref{BAK}, dynamics of the different polarization modes $\lambda = \{L,\pm\}$ decouple by taking into account orthogonality condition of polarization vectors: $\epsilon^{(L)}_i \epsilon_i^{\pm} = \hat{k}_i \epsilon^{(\pm)}_i = 0$. In particular, the equations of motion for the longitudinal and the transverse mode can be derived in terms of the cosmic time as 
\begin{align}\label{ALandAtCT}
\nn \ddot{A}_L + \frac{[3+2 \partial_t m_{\rm eff} / (H m_{\rm eff})]k^2 + a^2 m_{\rm eff}^2}{k^2 + a^2 m_{\rm eff}^2}\, H\dot{A}_L
 + \left(1 - \frac{(1+w)\, \alpha^2 H^2}{m_{\rm eff}^2}\right)\left(\frac{k^2}{a^2} + m_{\rm eff}^2\right)A_L &= 0, \\
 \ddot{A}_{\pm} + H \dot{A}_{\pm} + \left(\fr{k^2}{a^2} + (m_{\rm eff}^2 - (1+w)\, \alpha^2 H^2) \right) A_\pm &= 0.
\end{align}
Using \eqref{ALandAtCT}, we present an analytical understanding of mode evolution during the cosmic history in sections \ref{s2p1} and \ref{s2p3}.

\section{Mode evolution of $A_L$ during inflation: Numerical Procedure}\label{AppB}

In order to accurately capture the cosmological evolution of the individual $A_L$ modes in the post-inflationary universe, we need to determine their initial conditions at the beginning of RDU when they are outside the co-moving horizon. For this purpose, we solve for the transfer function $T_L$ \eqref{Alrdu} using the EoM \eqref{ALandAtinf} during inflation. Working with time variable $y = a/a_*$ and noting $ x \equiv - k \tau = - k \tau_* (\tau / \tau_*) = x_* y^{-1}$ and $x_* \equiv k/k_*$, the dynamics of the transfer function during inflation can be characterized by the following equation:
\begin{align}\label{Tfinf}
 T_L'' (y,x_*) &+ \left[1 + \frac{x_*^2}{x_*^2 + y^2 (1 + (\alpha^2 H_{\rm I}^2 / m^2))}\right] \fr{2}{y}\,\, T'_L (y,x_*)\\ \nn
&\quad\quad\quad\quad\quad\quad\quad\quad\quad\quad\quad\quad\quad\quad\quad +  \frac{m^2}{H_{\rm I}^2}\left(\frac{x_*^2}{y^4} + \frac{1 + (\alpha^2 H_{\rm I}^2 / m^2)}{y^2}\right)T_L(y,x_*) = 0.
\end{align}
Focusing on sizeable non-minimal coupling regime $\alpha H_{\rm I} / m \gg 1$, we work with sufficiently light longitudinal modes where $m_{\rm eff,I}^2 = m^2 + \alpha^2 H_{\rm I}^2 \simeq \alpha^2 H_{\rm I}^2 \ll H_{\rm I}^2$ implying the strength of the non-minimal coupling to be small, $\alpha \ll 1$. In this case, we can initialize the modes at horizon exit by safely assuming they are in the relativistic regime
parametrized by the solution \eqref{LRS}. In terms of the new time variable $y$, we therefore adopt the following initial conditions during inflation,
\beq\label{TLinfin}
T_L (y_{\rm in}, x_*) = (1 - i)\, {\rm e}^{i}, \quad\quad T_L' (y_{\rm in}, x_*) = - (x_*\, y_{\rm end}^2)^{-1} \, {\rm e}^{i},
\eeq
where for each mode we have $y_{\rm in} = x_* y_{\rm end}^2$ with $y_{\rm end} = \sqrt{m/H_{\rm I}}$ denoting the end of inflation/beginning of RDU. To determine accurate final conditions that can be used for the post-inflationary evolution of the longitudinal modes, we solve \eqref{Tfinf} for a grid of $x_* = k/k_*$ values using \eqref{TLinfin} for various $\alpha$ and $m / H_{\rm I} \ll 1$ choices. As we mentioned before,  we do so in a regime where effects introduced by the non-minimal coupling is sufficiently large during inflation, \ie
\beq
\alpha \frac{H_{\rm I}}{m} \gg 1 \quad \Longrightarrow\quad \frac{m}{H_{\rm I}} \ll \alpha \ll 1,
\eeq
where the final inequality is required to keep the longitudinal mode light during inflation. This procedure provides us with a pair of $\{ T_L(y_{\rm end}, x_*), T_L'(y_{\rm end}, x_*)\}$ for all $x_* = k/k_*$ values that can be used as an input for the post-inflationary evolution we discuss in Section \ref{Numrdu}. 
\section{Energy density of the longitudinal modes}\label{AppC}
In this appendix, we focus our attention to the energy density and pressure of the vector fields. For this purpose, we note that the energy-momentum tensor (EMT) of the action \eqref{sgf} can be written as:
\beq\label{tal}
T_{\mu\nu} \equiv - \fr{2}{\sqrt{-g}}\fr{\delta S_{\rm GF}}{\delta g^{\mu\nu}} = T^{(\alpha = 0 )}_{\mu\nu} + T^{(\alpha)}_{\mu\nu},
\eeq
where the first term is the well known EMT of the Proca action:
\beq\label{tpr}
 T^{(\alpha = 0 )}_{\mu\nu} = g^{\rho \sigma} F_{\mu\rho} F_{\nu \sigma} + m^2 A_\mu A_\nu - g_{\mu\nu} \left(\fr{1}{4}F_{\alpha\beta} F^{\alpha\beta} +\fr{1}{2}m^2 A_\beta A^{\beta}\right).
\eeq
It is quite non-trivial to obtain an explicit expression for $T^{(\alpha)}_{\mu\nu}$. Below, we present a detailed derivation of this contribution by explicitly varying terms $\propto \alpha^2$ in the action \eqref{sgf}:
\beq\label{varSlamb}
\delta S^{(\alpha)} = \fr{\alpha^2}{6} \int \d^4 x \sqrt{-g}\, \bigg[-\fr{1}{2}\,g_{\mu\nu}\, G_{\rho\sigma}\, A^{\sigma} A^{\rho}\,\, \delta g^{\mu\nu} + \delta G^{\mu\nu}\, A_\mu A_\nu\bigg],
\eeq
where we used $\delta \sqrt{-g} = - \sqrt{-g}\, g_{\mu\nu}\, \delta g^{\mu\nu}/2$ in the first term above. Utilizing the definition of the Ricci tensor with upper indices $R^{\mu\nu} = g^{\mu\alpha}\,g^{\nu\beta} R_{\alpha\beta}$ and the variation of the Ricci scalar $\delta R = R_{\rho\sigma} \delta g^{\rho\sigma} + g_{\rho\sigma}\, \delta R^{\rho\sigma}$, the variation of the second term of \eqref{varSlamb} can be written as 
\begin{align}\label{varET}
\delta G^{\mu\nu} = g^{\mu\alpha}g^{\nu\beta}\delta R_{\alpha\beta} - \fr{1}{2}\, g^{\mu\nu}\, g^{\rho\sigma}\, \delta R_{\rho\sigma} \,-\,& \fr{1}{2} \left(R\,\delta g^{\mu\nu}  + g^{\mu\nu} R_{\rho\sigma}\, \delta g^{\rho\sigma}\right) \\ \nn
& \quad\quad\quad\quad\quad\quad\quad + g^{\nu\beta} R_{\alpha \beta}\, \delta g^{\mu\alpha} + g^{\mu\alpha} R_{\alpha \beta}\, \delta g^{\nu\beta}.
\end{align}
Notice that the last terms in \eqref{varET} are already given as a variation of the inverse metric, so we only need to obtain the first two using the variation of the Ricci tensor in terms of the variation of the inverse metric \cite{Carroll:2004st}:
\begin{align}\label{varR}
\delta R_{\mu \nu} &= \,\nabla_\rho(\delta \Gamma^{\rho}_{\,\nu\mu})- \nabla_\nu(\delta \Gamma^{\rho}_{\,\rho\mu}),\\
\nn \delta \Gamma^{\sigma}_{\,\mu\nu} &= -\fr{1}{2} \left[\,g_{\lambda\mu}\nabla_{\nu}(\delta g^{\lambda\sigma}) + g_{\lambda \nu} \nabla_{\mu}(\delta g^{\lambda\sigma}) - g_{\mu\alpha}\,g_{\nu\beta}\nabla^{\sigma}(\delta g^{\alpha\beta})\,\right].
\end{align} 
We label these contributions in the the action \eqref{varSlamb} as $\delta S_{1}^{(\alpha)}$ and  $\delta S_{2}^{(\alpha)}$. Plugging \eqref{varR} and performing a couple of integration by parts, for the former we obtain
\begin{align}\label{rt1}
 \delta S_{1}^{(\alpha)} &= \fr{\alpha^2}{12} \int \d^4 x \sqrt{-g}\, \bigg[\nabla_\nu\nabla_\rho (A^{\rho}A_\mu)-\nabla_\rho\nabla_\nu (A^{\rho}A_\mu) -\nabla_\mu\nabla_\rho (A_\nu A^{\rho}) + \nabla_\rho\nabla_\mu (A_{\nu}A^\rho)\\ \nn 
&\quad\quad\quad\quad\quad\quad\quad\quad-\nabla_\rho\nabla_\mu (A_{\nu}A^\rho) - \nabla_\rho\nabla_\nu (A^\rho A_{\mu}) + \nabla^\rho\nabla_\rho (A_\mu A_{\nu}) + g_{\mu\nu} \nabla_\rho\nabla_\sigma (A^\rho A^{\sigma})\bigg]\delta g^{\mu\nu},
\end{align}
where for convenience we added and subtracted a term in passing from first to the second line. Using the product rule and the definition of the Riemann \eqref{Rdef} and  
Ricci tensor \eqref{Riccidef}, the first line of \eqref{rt1} can be recast as a term without derivatives on the gauge fields:
\begin{align}\label{flb6}
\textbf{\rm 1st line of (C.6)} &= -\nabla_\rho\nabla_\nu (A^{\rho}A_\mu) + \nabla_\nu\nabla_\rho (A^{\rho}A_\mu) -\nabla_\mu\nabla_\rho (A_\nu A^{\rho}) + \nabla_\rho\nabla_\mu (A_{\nu}A^\rho),\\\nn
& = - A_\mu ([\nabla_\rho,\nabla_\nu] A^{\rho}) - A^\rho ([\nabla_\rho,\nabla_\nu] A_\mu) + A_\nu ([\nabla_\rho,\nabla_\mu] A^{\rho}) + A^\rho ([\nabla_\rho,\nabla_\mu] A_\nu),\\\nn
& = - R_{\sigma\nu} A_\mu A^{\sigma} - R_{\mu\sigma\rho\nu} A^{\sigma} A^{\rho} + R_{\sigma\mu} A_\nu A^{\sigma} + R_{\nu\sigma\rho\mu} A^{\sigma} A^{\rho},\\ \nn
& = - R_{\sigma\nu} A_\mu A^{\sigma} + R_{\sigma\mu} A_\nu A^{\sigma} - R_{\mu\nu\rho\sigma} A^{\sigma} A^{\rho},\\ \nn 
& = - R_{\sigma\nu} A_\mu A^{\sigma} + R_{\sigma\mu} A_\nu A^{\sigma},
\end{align}
where in passing from second to third and from third to last line we have used the symmetry properties of the Riemann tensor. 

To derive the contribution to $T^{(\alpha)}_{\mu\nu}$ from the second term proportional to the variation of the Ricci tensor in \eqref{varET}, we again utilize \eqref{varR} and perform a couple integration by parts to obtain
\beq\label{rt2}
\delta S_{2}^{(\alpha)} = \fr{\alpha^2}{12} \int \d^4 x \sqrt{-g}\, \bigg[\nabla_\mu\nabla_\nu (A_\sigma A^{\sigma}) - g_{\mu\nu} \nabla^{\rho}\nabla_\rho (A_\sigma A^{\sigma})\bigg]\delta g^{\mu\nu}.
\eeq
 Compiling our findings in \eqref{varSlamb}, \eqref{varET}, \eqref{rt1}, \eqref{flb6} and \eqref{rt2}, the energy momentum tensor \eqref{tal} due to non-minimal coupling can be obtained as 
\begin{align}\label{tnm}
T^{(\alpha)}_{\mu\nu} &= \fr{\alpha^2}{6}\bigg[g_{\mu\nu}\, G_{\rho\sigma} A^{\sigma} A^{\rho}  + R_{\mu\nu} (A_\sigma A^{\sigma}) + R\, A_\mu A_\nu -3 R_{\sigma\nu}\, A_\mu A^{\sigma} - R_{\sigma\mu}\, A_\nu A^{\sigma}\\ \nn
&\quad\,\,\,+\big(\left[g_{\mu\nu}\,g^{\rho\sigma}-\delta^{\rho}_{\,\mu}\delta^{\sigma}_{\,\nu}\right]g^{\alpha\beta}+\delta^{\alpha}_{\,\nu}\delta^{\sigma}_{\,\mu}\,g^{\beta\rho} + \delta^{\beta}_{\,\mu}\delta^{\sigma}_{\,\nu}\,g^{\alpha\rho}-\delta^{\alpha}_{\,\nu}\delta^{\beta}_{\,\mu}\,g^{\sigma\rho}-g_{\mu\nu}\,g^{\alpha\rho}\,g^{\beta\sigma}\big)\nabla_{\rho}\nabla_{\sigma}(A_{\alpha} A_{\beta})\bigg],
\end{align}
where 
\beq\label{hdt}
\nabla_\rho \nabla_\sigma(A_\alpha A_\beta) = A_\beta \nabla_\rho (\nabla_\sigma A_{\alpha}) +  A_\alpha \nabla_\rho (\nabla_\sigma A_{\beta}) + \nabla_\rho A_\alpha \nabla_\sigma A_\beta + \nabla_\sigma A_\alpha \nabla_\rho A_\beta.
\eeq
Noting that $T_{00} = \rho$, we first expand the non-derivative terms explicitly using the components of Einstein and Ricci tensor in the first line of \eqref{tnm} together with the higher derivative terms in the second line in terms of the temporal and spatial component of the vector field as 
\beq\label{rhoa}
\rho^{(\alpha)}_{A} = \fr{\alpha^2}{6}\left[3H^2\left(\fr{1+w}{2}\right)\left(3A_0^2 +\fr{A_iA_i}{a^2}\right)+g^{ij}\left[\nabla_j,\nabla_0\right](A_0 A_i) + \left(g^{li}g^{kj}-g^{ij}g^{kl}\right)\nabla_i \nabla_j(A_l A_k)\right].
\eeq
Utilizing the definition of the Riemann tensor \eqref{Rdef}, the second term above read as 
\beq
g^{ij}\left[\nabla_j,\nabla_0\right](A_0 A_i) = - 3H^2 \left(\fr{1 + 3w}{6}\right)\left(3A_0^2 + \fr{A_i A_i}{a^2}\right).
\eeq
Finally unpacking the last term in \eqref{rhoa} according to the definition \eqref{hdt} and after a bit of algebra, we compile our findings to obtain the energy density due to non-minimal coupling as 
\begin{align}\label{rhoAa}
\nn \rho^{(\alpha)}_{A} &= \fr{\alpha^2}{6} \bigg[9H^2 A_0^2 - H^2 \fr{A_iA_i}{a^2} -\fr{(\vec{\nabla}\times \vec{A})^2}{a^4} - \fr{(\partial_i A_k)^2}{a^4} + \fr{\partial_i A_i \partial_j A_j}{a^4} + 2 \fr{A_i \partial_i \partial_k A_k}{a^4} - 2 \fr{A_i \vec{\nabla}^2 A_i}{a^4}\\ 
&\quad\quad\quad - 4H \fr{\partial_i (A_0 A_i)}{a^2} + 4H \fr{\dot{A}_k A_k}{a^2}\bigg].
\end{align}
On the other hand, energy density in the absence of non-minimal coupling can be obtained from \eqref{tpr} as
\beq\label{rhoAaz}
\rho^{(\alpha = 0)}_{A} = \fr{1}{2 a^2}\left[ (\dot{A}_i - \partial_i A_0)^2 + \fr{(\vec{\nabla} \times A)^2}{a^2} +m^2 (a^2 A_0^2 + A_i A_i)\right].
\eeq
To obtain the total energy density of the longitudinal modes from \eqref{rhoAa} and \eqref{rhoAaz}, we first notice that the curly of the vector field $\vec{\nabla} \times A$ does not contain the longitudinal mode  which is a direct consequence of the property of the polarization vector $\epsilon_i^{(L)}$ noted in the second line of \eqref{pve}. On the other hand going into the Fourier space using the eqs. \eqref{b4}-\eqref{conf}, one can show that the combination of the last four terms in the first line, as well as the first term in the last line of \eqref{rhoAa} vanishes. Furthermore, as far as the longitudinal mode is concerned, the envelope of the last term in the non-minimal EMT behaves as $\sim H \dot{A}_k A_k \sim H^2 A_k A_k$ at late times, as can be inferred from eq. \eqref{ALNR}. Therefore, in the late time limit the energy density of the longitudinal modes that stem from the non-minimal coupling behave as
\beq
\rho^{(\alpha)}_{A_L} \simeq \fr{1}{2a^2} \left[3 \alpha^2 H^2 a^2 A_0^2 + H^2 A_k A_k \right].
\eeq
Comparing the overall scaling of these terms with the last terms in \eqref{rhoAaz}, we conclude that the corrections to the energy density due to non-minimal interactions are much less important in the late time limit at which we are interested to evaluate the abundance of the longitudinal modes. In particular, for $a \gg a_* \gg a_{\rm c}$, we have $m \gg \alpha H$  because $\alpha H$ is already comparable to $m$ at the end of the phase {\bf (G)} in the RDU (see Fig. \ref{fig:SAL2}). Therefore, we can safely adopt the standard energy density \eqref{rhoAaz} in order to compute the late time abundance of the longitudinal modes. Using the Fourier decomposition \eqref{b4},\eqref{b6} and late time version of the constraint \eqref{conf} (\eg $m_{\rm eff}\to m$):
\beq\label{confl}
A_0(t,\vbf{k}) = - i\fr{\, ({k}_i\,\epsilon^{(L)}_i(\vbf{k}))}{k^2 + a^2 m^2}\,\dot{A}_L(t,\vbf{k}),
\eeq
the expectation value of $\rho_{A_L}$ \eqref{rhoAaz} can thus be written as
\begin{align}\label{rhoal}
\nn \langle \rho_{A_L}\rangle = \fr{1}{2a^4}\int \fr{\d^3 k\, \d^3 k'}{(2\pi)^3}e^{i(\vbf{k}+\vbf{k}').\vbf{x}}& \bigg\{\bigg[\epsilon^{(0)}_i(\vbf{k}) \epsilon^{(0)}_i(\vbf{k}')- 2 \fr{(k'_i \epsilon_i(\vbf{k}))\, (k'_i \epsilon^{(0)}_i(\vbf{k}'))}{k^2 + a^2 m^2} \\\nn &\quad+\fr{(k_ik_i'-a^2m^2)\,(k_i \epsilon^{(0)}_i(\vbf{k}))(k_j' \epsilon^{(0)}_j(\vbf{k}'))}{(k^2 + a^2 m^2)\,(k'^{2} + a^2 m^2)}\bigg]\langle A_L' (\tau,\vbf{k})A_L'(\tau,\vbf{k}')\rangle\\
&\,\,+ a^2 m^2\, \epsilon^{(0)}_i(\vbf{k}) \epsilon^{(0)}_i(\vbf{k}')\, \langle A_L (\tau,\vbf{k})A_L(\tau,\vbf{k}')\rangle  \bigg\},
\end{align}
where we switched to conformal time. To evaluate the expectation values inside the integrand, we follow the canonical quantization procedure of the longitudinal mode via
\beq
A_L (\tau, \vbf{k})\to \hat{A}_L (\tau,\vbf{k}) = A_L(\tau,k)\, \hat{a}_L(\vbf{k}) + A_L^{*}(\tau,k)\, \hat{a}_L^{\dagger}(-\vbf{k})
\eeq
where the annihilation and creating operators satisfy $[\hat{a}_L(\vbf{k}),\hat{a}_L^{\dagger}(\vbf{k}')] = \delta(\vbf{k}-\vbf{k}')$. Then defining the power spectrum of a hermitian operator $\hat{X}$ as 
\beq
\langle \hat{X}(\tau,\vbf{k}) \hat{X}(\tau,\vbf{k}')\rangle = \delta (\vbf{k}+\vbf{k}') \fr{2\pi^2}{k^3} \mathcal{P}_X (\tau,k),
\eeq
the expectation value of the energy density \eqref{rhoal} becomes
\beq
\langle \rho_{A_L}\rangle = \fr{1}{2a^4}\int \d \ln k\, \bigg\{\fr{a^2 m^2}{k^2 +a^2 m^2} \,\mathcal{P}_{A'_L}(\tau,k) + a^2 m^2\, \mathcal{P}_{A_L}(\tau,k)\bigg\}.
\eeq

\section{Abundance of transverse vector modes}
\label{sec_trab}
In the main text we ignored the contribution to the abundance of the vector field from the transverse modes, relying on the fact that they stay in their vacuum configuration during inflation in the small mass limit, $m/H_{\rm I} \to 0$. However, post-inflationary evolution of the transverse modes could  potentially provide significant contribution to the abundance since they obtain a tachyonic mass following inflationary phase in the RDU era for the sizeable non-minimal coupling regime $\alpha H_{\rm I}/m \gg 1$ we are operating. This could be seen from eq. \eqref{ALandAt}, where the mass of the transverse mode in the post-inflationary era behave as 
\beq\label{mT}
m_{T}^2 =
 \begin{dcases} 
       m^2 - \frac{\alpha^2 H_{\rm I}^2}{3}\left(\fr{a_{\rm end}}{a}\right)^4 \simeq  - \frac{\alpha^2 H_{\rm I}^2}{3}\left(\fr{a_{\rm end}}{a}\right)^4,& \quad a_{\rm end} \leq a \leq a_{\rm tc} \,,\\
        m^2 & \quad\quad\quad\quad a > a_{\rm tc},
   \end{dcases}
\eeq
where the end of tachyonic regime with respect to the end of inflation can be identified as 
\beq
\fr{a_{\rm tc}}{{a_{\rm end}}} = 3^{-1/4} \sqrt{\frac{\alpha H_{\rm I}}{m}} > 1. 
\eeq
Notice that the end of the tachyonic era almost coincides with the end of phase {\bf (G)} \eqref{DG} during which the longitudinal modes have gradient type instability. This similarity notwithstanding, the transverse modes that are affected by the tachyonic instability are different as compared to the former. To understand this, we rewrite the the mode equation of the transverse modes in terms of the variables we are familiar from the the previous sections as
\beq\label{EoMATRDU}
A_{\pm}''(y,x_*) + \left(x_*^2 + y^2 \left[1 - \fr{\alpha^2 H^2}{3 m^2}\right]\right)A_{\pm}(y,x_*) = 0,
\eeq
where we again note $H(y) = H_{\rm I}\, (y_{\rm end}/y)^2$ during RDU.
From the effective dispersion relation of the expression above, we infer that the shortest modes that could be affected by the presence of the tachyonic regime satisfy
\beq
x_* < \frac{1}{\sqrt{3}} \frac{\alpha}{y}\quad \Longrightarrow\quad  x_* < \frac{1}{\sqrt{3}} \frac{\alpha}{y_{\rm end}} = \frac{\alpha}{\sqrt{3}}\sqrt{\frac{H_{\rm I}}{m}} \ll 1.
\eeq
Therefore, as compared to the longitudinal modes, relatively large scale transverse modes could be potentially amplified during the tachyonic regime $a_{\rm end} < a < a_{\rm tc}$.

To understand if the transverse modes can be amplified in the tachyonic era, we followed a similar approach we carry in Appendix \ref{AppB} to first generate initial conditions for the modes in the post-inflationary era by evolving $A_T$ during inflation from deep inside the horizon until the end of inflation. We then use these initial condition to numerically evolve the mode equation of the transverse modes \eqref{EoMATRDU} during RDU until sufficiently late times. In this way, we found that transverse modes that could potentially exhibit instability do not behave any interesting way during the tachyonic era as compared with the vanishing non-minimal coupling case $\alpha \to 0$ for which the tachyonic phase is absent. We will not present these null results here, however we refer the interested to the {\sf Mathematica} \href{https://github.com/oozsoy/NM_Vector-Field_CosmoEval/blob/main/README.md}{notebook} that lead us draw this conclusion. As can be inferred from the numerical analysis we provide, the absence of enhancement can be understood by 
\begin{itemize}
\item the fact that transverse modes stay in their vacuum configuration (recall the discussion in section \ref{s2p1}) during the inflationary era, reaching the beginning of radiation dominated phase, with very small velocities $|\sqrt{2k} A'(y,x_*)|\approx x_*$ for the potentially tachyonic modes that satisfy $x_* \ll \ll 1$.
\item the fact that tachyonic era is short and transient and fails to generate sizeable velocity gradients $A_T'(y,x_*)$ that can trigger amplification for the transverse modes, in particular for the parameter space that realizes interesting abundance of longitudinal modes as dark matter.
\end{itemize}
On the other hand, outside the tachyonic region, particle production solely due expansion of the universe typically leads to small energy density in the transverse modes as compared to the longitudinal modes \cite{Kolb:2017jvz,Ema:2019yrd,Kolb:2020fwh}. Therefore, we expect that transverse modes do not play an important role in our proposal and the final abundance of  dark matter is set by the longitudinal modes parametrized by the expression \eqref{ALA}. 

\end{appendix}

\addcontentsline{toc}{section}{References}
\bibliographystyle{utphys}
\bibliography{paper2}
\end{document}